\documentclass[conference,twoside]{IEEEtran}

\usepackage[utf8]{inputenc}
\usepackage[T1]{fontenc}
\usepackage{amsmath}
\usepackage{graphicx}
\usepackage{caption}
\usepackage{subcaption}
\usepackage{booktabs}
\usepackage{comment}
\usepackage{tabularx}
\usepackage{array}
\usepackage{makecell}
\usepackage{fancyhdr}
\usepackage{xcolor}

\title{High-Performance Rotor Cooling with Ducted Liquid in Completely Cold-Formed Modular Motor Shaft}

\author{
    \IEEEauthorblockN{Rezvan Alamian, Sören Müller, Uwe Steinmetz, Christian Henrich, Stefan Götz}
}

\begin{document}

\maketitle

\pagestyle{fancy}
\fancyhf{}  

\fancyhead[RO]{\textcolor{black}{\fontsize{9}{11}\selectfont High-Performance Rotor Cooling \ldots}}
\fancyhead[LE]{\textcolor{black}{\fontsize{9}{11}\selectfont Alamian et al., 2025}}
\fancyhead[RE,LO]{\thepage}  

\renewcommand{\headrulewidth}{0pt}

\fancypagestyle{firstpage}{
    \fancyhf{}  
    \fancyhead[R]{\textcolor{black}{\fontsize{9}{11}\selectfont }}
    \fancyhead[L]{\thepage}
    \renewcommand{\headrulewidth}{0pt}
}
\thispagestyle{firstpage}

\begin{abstract}
This study suggests a novel rotor-cooling shaft concept for high-performance electric motors that increases the effectiveness of cooling and is yet simple and cost-effective to manufacture. In that context, we investigate the thermal performance of four shaft geometries for rotor cooling in automotive applications. The proposed tooth-guided liquid-cooling shaft design aims to solve the high churning loss of conventional cooled rotor shafts due to internal vortex formation and their still limited heat transfer. Therefore, we  optimize heat transfer efficiency and pressure management by incorporating cold-formed internal channels that restrict vortex formation beyond a degree that improves heat transfer. We evaluated key performance metrics, including heat transfer rate, outlet temperature, pressure drop, and velocity profiles, under varying rotational speeds, inlet flow rates, and coolant temperatures. Computational fluid analysis demonstrates that the tooth-guided design outperforms conventional hollow shafts and achieves up to 110\% higher cooling efficiency at low rotational speeds, while it maintains comparable pressure levels. These findings provide practical insight into geometry-driven thermal optimization and offer a path toward improving the performance and durability of electric motors.
\end{abstract}


\section{Introduction}
\IEEEPARstart{T}{he} rapid advancement of electric vehicles (EVs) has stimulated significant improvements in electric motor design, particularly in the areas of thermal management and efficiency \cite{Xu2023, GLRReview}. Electric motors generate substantial heat due to electromagnetic and mechanical losses, which, if not effectively managed, can limit output power, reduce efficiency, and even cause premature component failure \cite{Agamloh2020}. This heat poses a critical challenge as it directly affects the reliability and lifespan of EV powertrains. Furthermore, the achievable power density of motors is largely determined by how much heat can be extracted by proper cooling. The power density of motors in electric vehicles has increased by more than an order of magnitude in the last decade due to more direct cooling. So far, most cooling technologies have focussed at the stator and the stator winding, though the rotor often contains one of the most heat-sensitive elements (magnetics in permanent-magnet synchronous machines) or can generate substantial heating (squirrel-cage induction machines).

Forced air cooling has traditionally been the primary method for managing heat in industrial and early automotive electric motors \cite{Gundabattini2022}. This approach uses fans or blowers to direct airflow over the motor's surface and relies on convective heat transfer to dissipate thermal energy. While this method is effective for cooling smaller motors or auxiliary components, it has been proven inadequate for high-performance EVs. The limited thermal conductivity and heat capacity of air result in insufficient cooling \cite{Gobbi2024}. 

Liquid stator cooling, particularly water jacket cooling, has become the dominant cooling system in electric vehicle motors \cite{Li2020}. Various types of cooling jackets, such as single-pass, multi-pass, and spiral designs, have been developed to enhance heat dissipation efficiency. A liquid coolant, typically water mixed with antifreeze agents, circulates through channels around the stator back, absorbs heat from the motor, and dissipates it through a radiator \cite{Jollyn2022}. Water-based coolants offer a high specific heat capacity, but other fluids such as oil and dielectric fluids are less corrosive and not electrically conductive to enable more direct contact with electrical components  \cite{Bourgault2019, Gronwald2021}.
However, the focus of all cooling on the stator can struggle to manage the heat generated within the rotor, which becomes critical in permanent synchronous machines at higher speeds with high magnet loss or at high torque and low speeds for induction machines.

%

In addition to active cooling of the stator and its winding, also the rotor is a thermally critical component. In permanent magnet synchronous machines, rare-earth magnets are thermally sensitive and limit the power particularly at high speeds when they heat up quickly due to eddy-current and similar high-frequency loss. In induction machines, the rotor often dominates the losses, particularly below the field-weakening regime at lower speeds, where the excitation induced into the squirrel cage can reach very high current levels. 
Thus, it was an induction motor that first established liquid rotor cooling in cars \cite{US9030063}.
At present, hollow shafts cool the rotors in several electric vehicles with induction machines \cite{US9030063, Audi} and are promising also for permanent-magnet synchronous machines \cite{Song2023, Gai2018, Gai2019, Wang2020}.
However, existing cooling shafts suffer from high churning losses in the coolant during rotation and still limited heat transfer, which requires an additional powerful oil pump. Furthermore, the coolant has to flow evenly throughout a wide range of rotor speeds as uneven cooling can lead to hot spots and potential failures \cite{Gai2020}.

\begin{figure}[htbp]
    \centering
    \begin{subfigure}{0.8\columnwidth} 
        \centering
        \includegraphics[width=\linewidth]{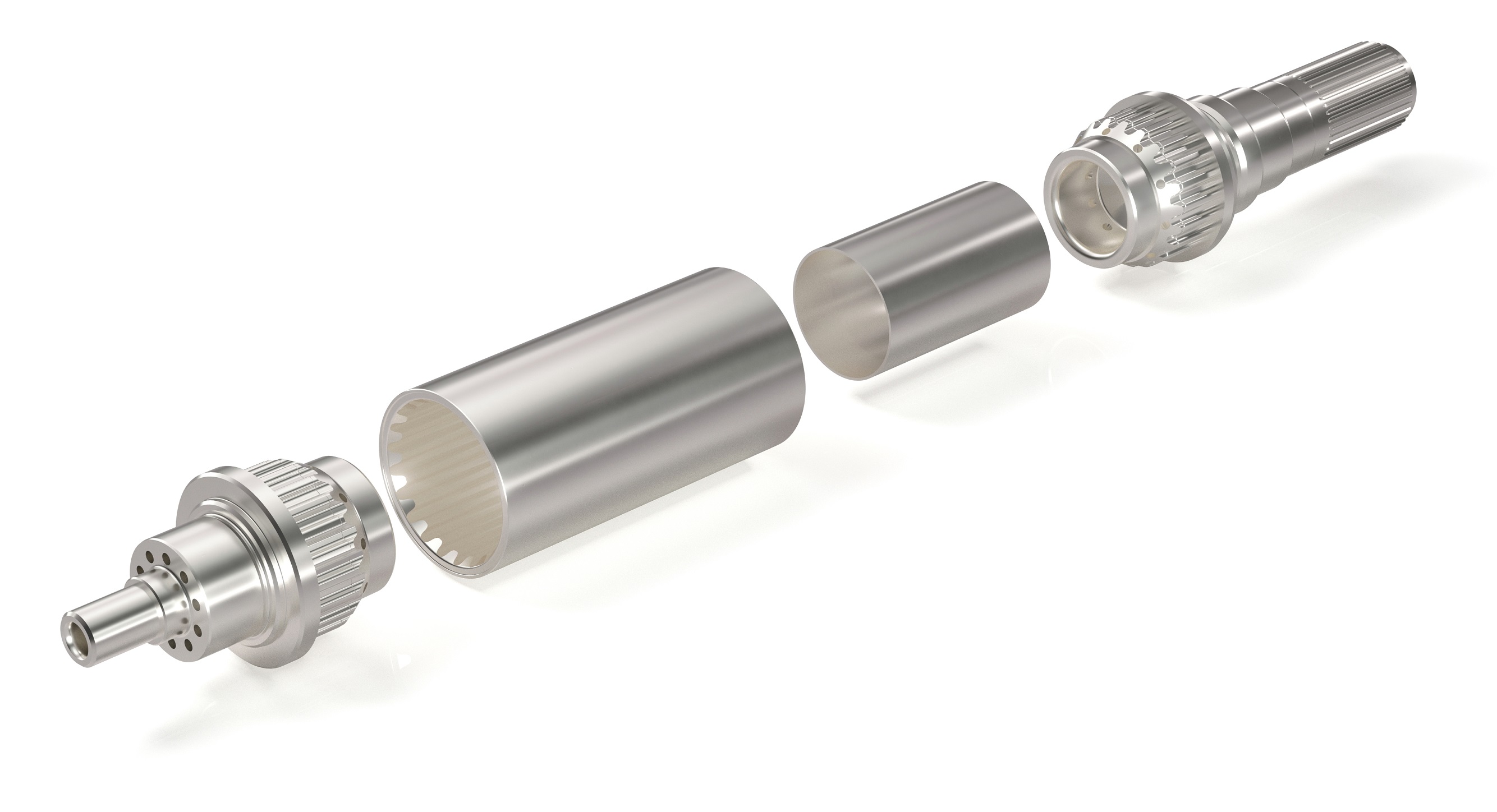} 
        \caption{Tooth-guided liquid-cooling shaft design.}
        \label{fig:quick1}
    \end{subfigure}
    \hfill 
    \begin{subfigure}{0.8\columnwidth}
        \centering
        \includegraphics[width=\linewidth]{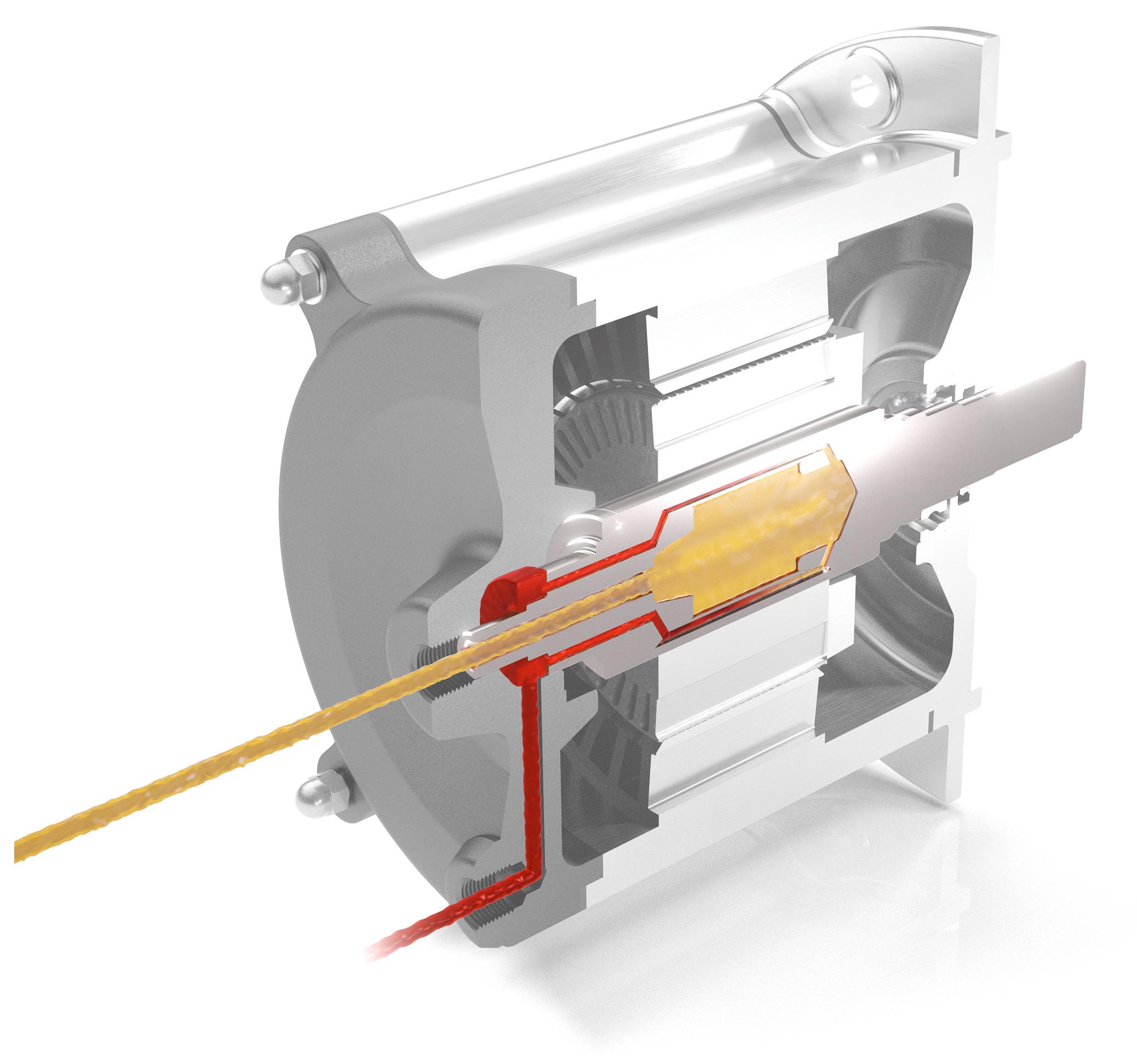}
        \caption{Integration of the tooth-guided liquid-cooling shaft within a motor.}
        \label{fig:quick2}
    \end{subfigure}
    \caption{Illustration of the tooth-guided liquid-cooling shaft design and its application in motor thermal management.}
    \label{fig:quickncool}
\end{figure}

We introduce a novel liquid-cooled shaft to optimize the thermal performance at low coolant pressure levels and low manufacturing cost, entirely in cold-forming technology. This design integrates guided liquid channels within the rotor shaft to significantly enhance the heat-transfer capabilities compared to conventional hollow shafts. Furthermore, the teeth suppress larger vortices, which turn out to be responsible for most of the churning losses in the state of the art. Figures~\ref{fig:quick1} and~\ref{fig:quick2} illustrate the  tooth-guided liquid-cooling shaft design and its integration into a motor assembly. The design incorporates several features in response to challenges of the state of the art. The shaft includes a tube-in-tube construction and an internal toothed profile, which optimize heat transfer by increasing the surface area in contact with the coolant. Additionally, the assembly process is streamlined by avoiding welding. Instead it uses entire cold-formed components and press-fit contacts for cost-effective series production. 

This design is scalable and has a variety of degrees of freedom, such as the size and shape of the teeth and the inner confinement, which affect the liquid stream, the surface area, and the vortex formation. This study therefore evaluates the performance compared to the state of the art, analyzes the reasons to stimulate further improvements of liquid rotor cooling, and study the impact of the key design parameters.
We focused on key performance indicators such as heat-transfer rates, pressure management, and temperature control under various operational conditions. Specifically, we examined the effects of varying coolant flow rates, shaft rotational speeds, and inlet temperatures on the thermal behavior of the tooth-guided liquid-cooling shaft. We compared temperature distributions along the shaft, analyzed the efficiency of heat dissipation, and evaluated the pressure drop across the coolant channels. Additionally, we explored how the internal toothed profile enhances overall cooling efficiency, particularly by improving heat transfer while maintaining stable pressure levels. By benchmarking these results against existing cooling methods, we highlight the potential advantages of the tooth-guided liquid-cooling shaft, especially for high-performance electric vehicle motors where effective thermal management is essential.

The structure of this paper is as follows: Section II describes the geometries of the four shaft models analyzed, including their dimensions and boundary conditions, along with material properties and mesh definitions. Section III outlines the methodology used in the study, covering the computational fluid dynamics (CFD) models, simulation setup, and convergence analysis performed to ensure reliable results. Section IV presents the results and analyzes the impact of rotational speed, inlet flow rate, and inlet temperature on the thermal and fluid dynamics performance of the shafts. Section V explains key design and manufacturing aspects of the proposed shaft. It covers material selection, cold-forming processes, and modular assembly techniques. These methods enable scalable and cost-effective production. Section VI discusses the implications of the thermal performance results for improving motor cooling systems in electric vehicles overall and highlights the potential advantages of the tooth-guided liquid-cooling shaft compared to conventional designs. Finally, Section VII concludes the key findings and suggests open questions as well as promising aspects for further improvements of cooling.

\section{Geometry Description}

\subsection{Geometries}
\begin{figure}[htbp]
\centering
\begin{subfigure}{0.5\textwidth}
    \centering
    \includegraphics[width=\textwidth]{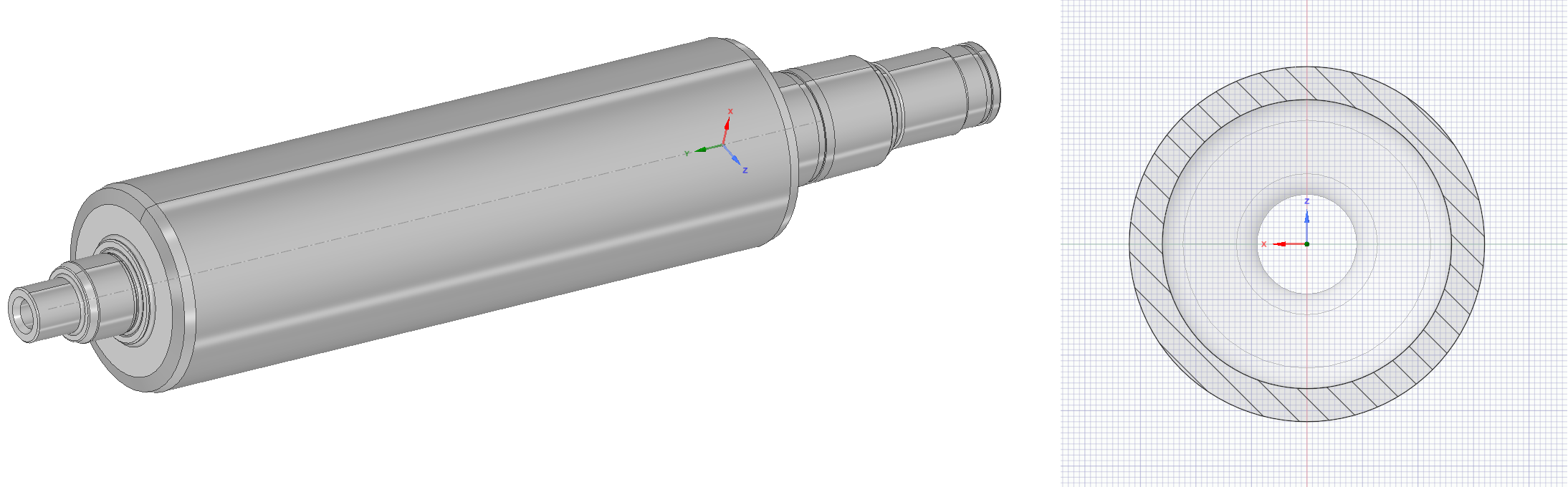}
    \caption{Shaft Model 1: Entirely hollow shaft.}
    \label{fig:geometry-model1}
\end{subfigure}
\begin{subfigure}{0.5\textwidth}
    \centering
    \includegraphics[width=\textwidth]{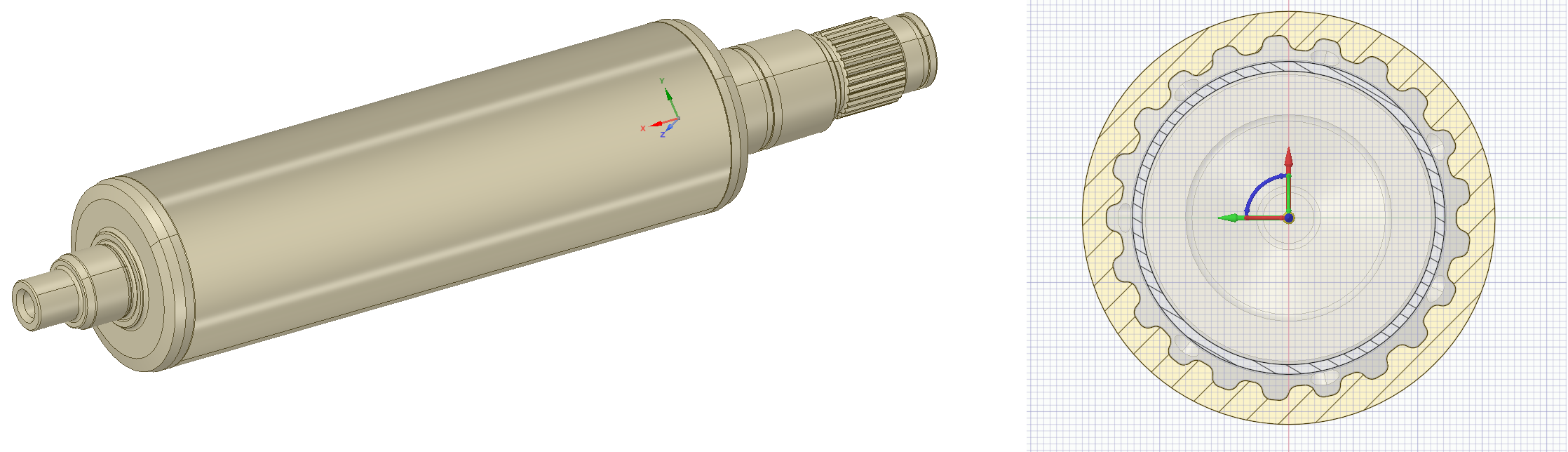}
    \caption{Shaft Model 2: Outside-toothed inner wall constrained by a smooth inner can.}
    \label{fig:geometry-model2}
\end{subfigure}
\begin{subfigure}{0.5\textwidth}
    \centering
    \includegraphics[width=\textwidth]{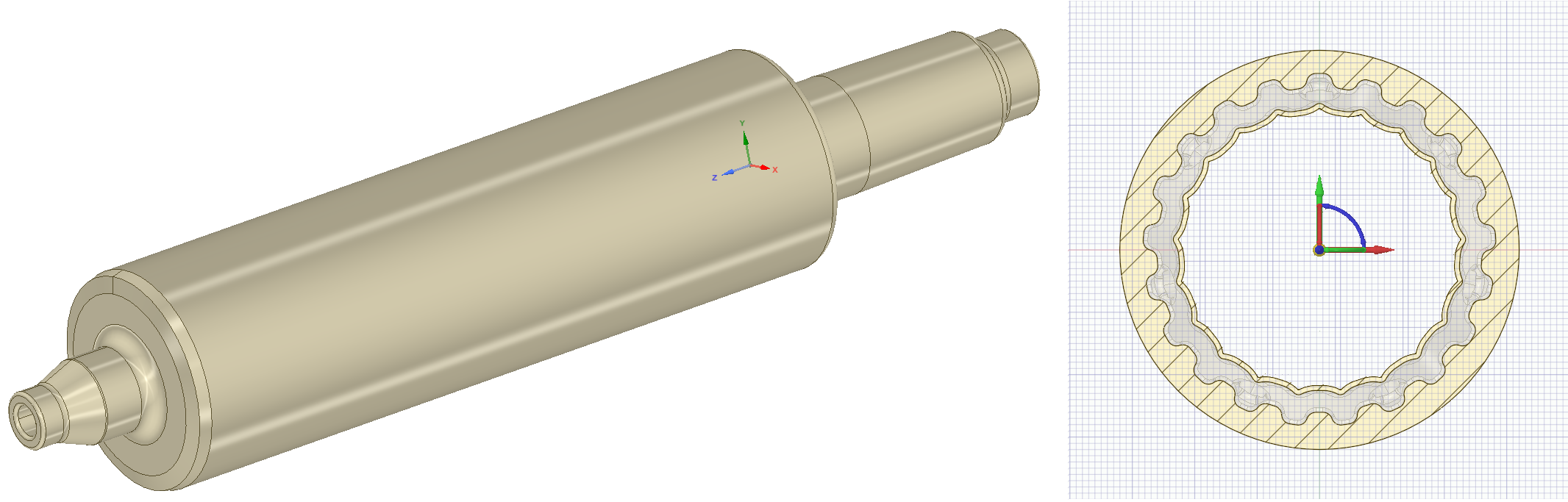}
    \caption{Shaft Model 3: Outside-toothed inner wall constrained by a wavy inner can.}
    \label{fig:geometry-model3}
\end{subfigure}
\begin{subfigure}{0.5\textwidth}
    \centering
    \includegraphics[width=\textwidth]{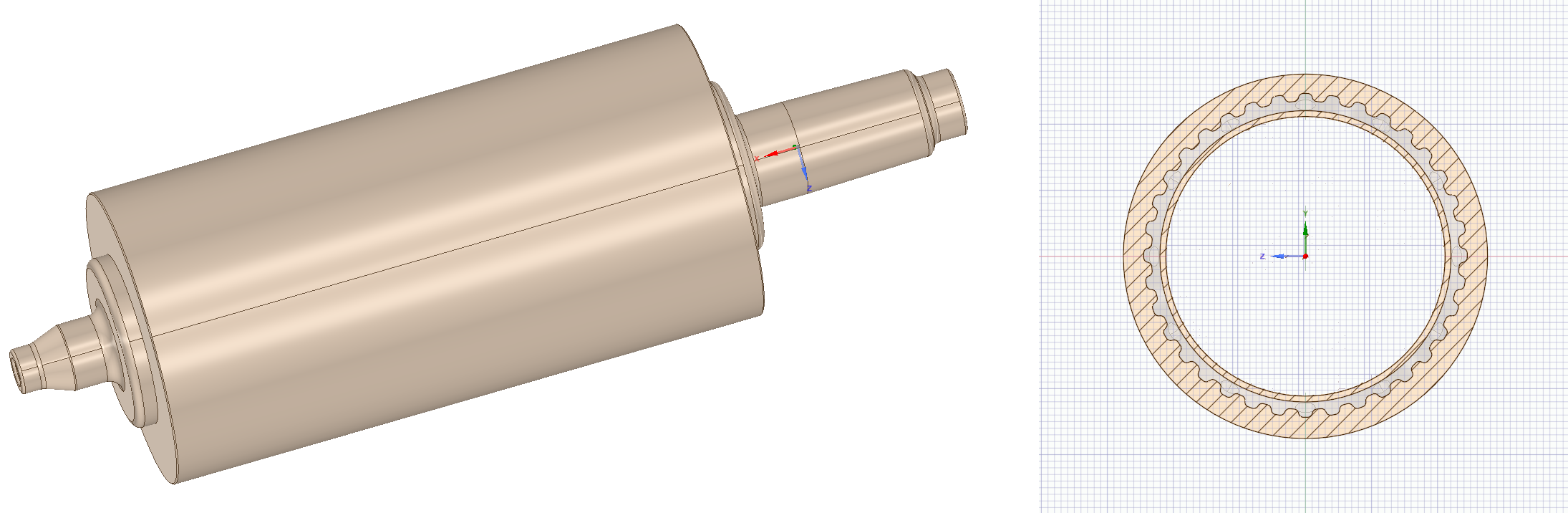}
    \caption{Shaft Model 4: Outside-toothed inner wall with more teeth constrained by a smooth inner can.}
    \label{fig:geometry-model4}
\end{subfigure}
\caption{Different shaft geometries.}
\label{fig:geometries}
\end{figure}

The study examines four distinct shaft geometries, each with unique design features aimed at improving cooling performance (Fig.\ \ref{fig:geometries}). The first geometry features a hollow cooling channel with a circular cross-section (\textit{Shaft Model 1}, Fig.\ \ref{fig:geometry-model1}), which serves as the benchmark for comparison. The second geometry (\textit{Shaft Model 2}, Fig.\ \ref{fig:geometry-model2}) incorporates a toothed channel near the shaft's outer surface to increase the cooling surface area. \textit{Shaft Model 3} (Fig.\ \ref{fig:geometry-model3}) includes toothed channels on the outer surface and a wavy inner profile to improve vortex control and pressure reduction. The fourth geometry (\textit{Shaft Model 4}, Fig.\ \ref{fig:geometry-model4}) resembles the second but with more channels and a larger outer diameter for increased cooling efficiency.

\subsection{Dimensions and Boundary Conditions}

\begin{figure}[htbp]
\centering
\includegraphics[width=0.45\textwidth]{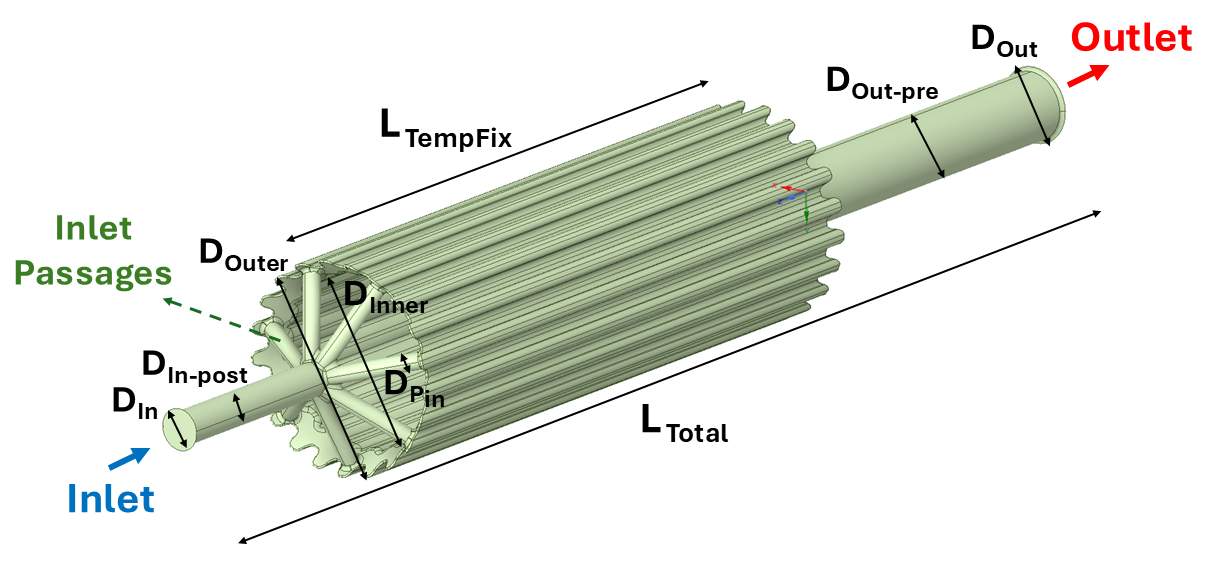}
\caption{\label{fig:rotor_shaft_dimensions} {\small Shaft specifications.}}
\end{figure}

Figure \ref{fig:rotor_shaft_dimensions} serves as a schematic representation of the various shaft designs and defines the key dimensions.
Table \ref{tab:rotor_shaft_dimensions} complements this figure with numerical values for four different shafts. The table also presents further specifications such as the fixed temperature area and interface area, which relate to the surface areas available for heat transfer.

\begin{table}[htbp]
    \centering
    \caption{Shaft Specifications}
    \label{tab:rotor_shaft_dimensions}
    \resizebox{0.45\textwidth}{!}{%
    \begin{tabular}{@{}l
      >{\centering\arraybackslash}p{2.5cm}
      >{\centering\arraybackslash}p{1.2cm}
      >{\centering\arraybackslash}p{1cm}
      >{\centering\arraybackslash}p{1cm}
      >{\centering\arraybackslash}p{1cm}
      >{\centering\arraybackslash}p{1cm}
      >{\centering\arraybackslash}p{1cm}
      >{\centering\arraybackslash}p{1cm}
        @{}}
      \toprule
      Shaft & \multicolumn{8}{c}{Dimensions (mm)} \\
      \cmidrule(r){2-9}
      & L\textsubscript{Total} & L\textsubscript{TempFix} & D\textsubscript{In} &
      D\textsubscript{In-post} & D\textsubscript{Out} & D\textsubscript{Out-pre} &
      D\textsubscript{Outer} & D\textsubscript{Inner} \\
      \midrule
      Shaft 1 & 340.35 & 187.3 & 11.3 & 9 & 18 & -- & 52.1 & -- \\
      Shaft 2 &  340.35 & 169.2 & 11.3 & 9 & 18 & 18 & 56.6 & 47.8 \\
      Shaft 3 &  340.35 & 166.6 & 11.0 & 9 & 20 & 18 & 56.6 & 45.6 \\
      Shaft 4 &  359.05 & 171.2 & 11.0 & 9 & 20 & 18 & 98.0 & 88.2 \\
      \midrule
      Shaft & \multicolumn{8}{c}{Additional Specs} \\
      \cmidrule(r){2-9}
      &  Fix Temp Area (mm\textsuperscript{2}) &  Interface Area (mm\textsuperscript{2}) & 
      No. of Inlet Passages & D\textsubscript{Pin} &  No. of Outlet Passages &
      D\textsubscript{Pout} &  No. of Teeth Channels & \\
      \midrule
      Shaft 1 &  30660 &  41961 & -- & -- & -- & -- & -- & \\
      Shaft 2 &  35476 &  47547 & 7 & 4.5 & 7 & 4.5 & 21 & \\
      Shaft 3 &  34968 & 46489 & 7 & 4.5 & 7 & 4.5 & 21 & \\
      Shaft 4 &  63882 & 90393 & 12 & 6 & 12 & 6 & 36 & \\
      \bottomrule
    \end{tabular}}
  \end{table}

\begin{figure}[htbp]
\centering
\includegraphics[width=0.5\textwidth]{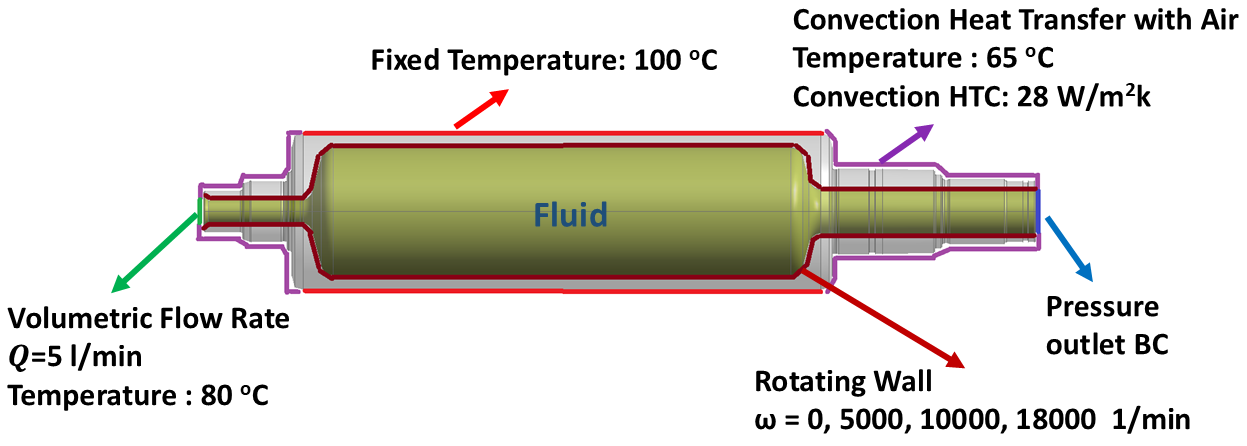}
\caption{\label{fig:geometryBCs_model1} {\small Schematic of the first shaft model along with the boundary conditions applied in ANSYS Fluent.}}
\end{figure}

We designed CFD representations for the base models in ANSYS Fluent. Figure~\ref{fig:geometryBCs_model1} illustrates the geometry of the first model and the boundary conditions. The boundary conditions for the first model are as follows:

\begin{itemize}
    \item \textbf{Inlet:}
    \begin{itemize}
        \item \textbf{Flow Rate:} $5 \, \textrm{l/min} = 8.3333 \times 10^{-5} \, \textrm{m}^3/\textrm{s}$.
        \item \textbf{Temperature:} $80~^\circ \textrm{C}$.
    \end{itemize}
    \item \textbf{Outlet:} Pressure outlet boundary condition.
    \item \textbf{Walls:} Solid wall surrounding the geometry, with the upper part set at $100~^\circ \textrm{C}$ and the rest exposed to air with free convection at $65~^\circ \textrm{C}$. The solid walls rotate at various rotational speeds of 0, 5000, 10000, and 18000~1/min.
\end{itemize}

\vspace{-10pt}

\begin{figure}[htbp]
\centering
\includegraphics[width=0.5\textwidth]{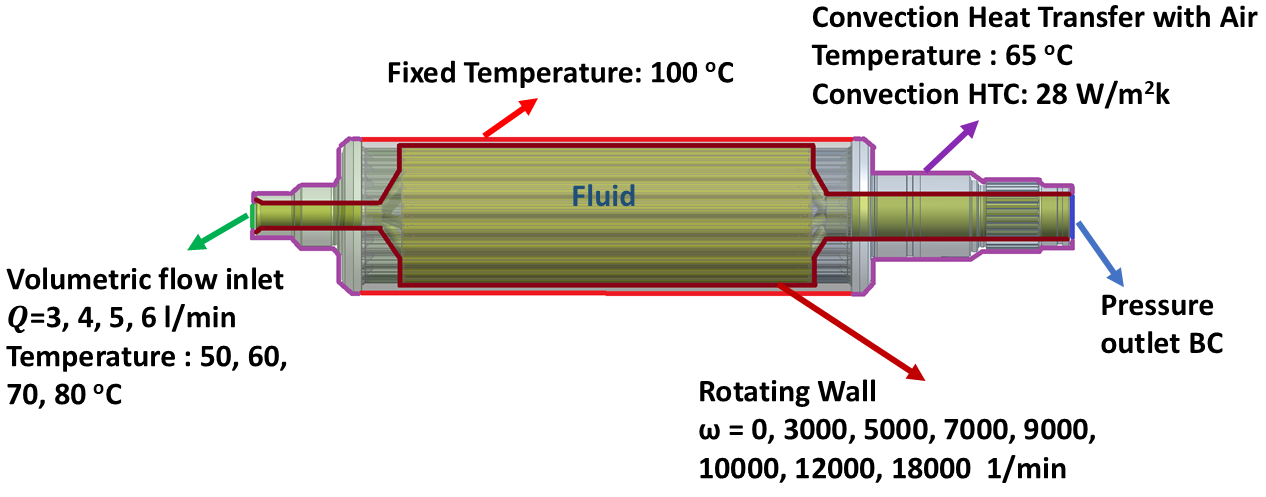}
\caption{\label{fig:geometryBCs_model2} {\small Schematic of the second shaft model along with the boundary conditions applied in ANSYS Fluent.}}
\end{figure}

Figure~\ref{fig:geometryBCs_model2} illustrates the geometry of the second model and the boundary conditions applied within ANSYS Fluent. While only the geometry of the second model is shown here to avoid repetition, the third and fourth models share the same boundary conditions. The boundary conditions for the second, third, and fourth models are as follows:

\begin{itemize}
    \item \textbf{Inlet:}
    \begin{itemize}
        \item \textbf{Flow Rate:} \(3, 4, 5, 6 \, \textrm{l/min}\).
        \item \textbf{Temperature:} \(50, 60, 70, \textrm{and}\ 80\,^\circ \textrm{C}\).
    \end{itemize}
    \item \textbf{Outlet:} Pressure outlet boundary condition.
    \item \textbf{Walls:} Solid wall surrounding the geometry, with the upper part set at \(100\,^\circ \textrm{C}\) and the rest exposed to air with free convection at \(65\,^\circ \textrm{C}\). The solid walls rotate at various rotational speeds of 0, 3000, 5000, 7000, 9000, 10000, 12000, and 18000~1/min.
\end{itemize}

The shaft material is 20MnCr5. Its temperature-dependent properties are detailed in Table~\ref{tab:rotor_shaft_properties}. The chosen cooling fluid is oil, specifically Fuchs FES 821-6436A ATF. Table~\ref{tab:oil_properties} lists its properties at various temperatures.

\begin{table}[htbp]
\centering
\small
\caption{Properties of the Rotor Shaft with Material 20MnCr5 at Different Temperatures}
\label{tab:rotor_shaft_properties}
\begin{tabularx}{\columnwidth}{|>{\centering\arraybackslash}p{0.12\columnwidth}|>{\centering\arraybackslash}p{0.14\columnwidth}|>{\centering\arraybackslash}p{0.2\columnwidth}|>{\centering\arraybackslash}p{0.19\columnwidth}|>{\centering\arraybackslash}p{0.17\columnwidth}|>{\centering\arraybackslash}p{0.17\columnwidth}|}
\hline
\makecell{Temp. \\ (°C)} & \makecell{Density \\ (Kg/m\textsuperscript{3})} &  \makecell{Thermal \\ Conductivity \\ (W/m$\cdot$K)} & \makecell{Thermal \\ Diffusivity \\ ($10^{-6}~\textrm{m}^2\textrm{/s}$)} & \makecell{Specific \\ Electrical \\ Resistance \\ (µ$\Omega$m)} & \makecell{Average \\ Specific \\ Heat \\ (kJ/kg$\cdot$K)} \\
\hline
20  & 7850 & 45.9 & 12.7 & 0.227 & 0.46 \\
100 & 7850 & 45.77 & 12.0 & 0.276 & 0.47 \\
200 & 7850 & 45.60 & 11.0 & 0.346 & 0.49 \\
\hline
\end{tabularx}
\end{table}

\begin{table}[htbp]
\centering
\small
\caption{Properties of the Oil: Fuchs FES 821-6436A ATF at Different Temperatures}
\label{tab:oil_properties}
\begin{tabularx}{\columnwidth}{|>{\centering\arraybackslash}p{0.11\columnwidth}|>{\centering\arraybackslash}p{0.13\columnwidth}|>{\centering\arraybackslash}p{0.18\columnwidth}|>{\centering\arraybackslash}p{0.16\columnwidth}|>{\centering\arraybackslash}p{0.21\columnwidth}|>{\centering\arraybackslash}p{0.2\columnwidth}|}
\hline
\makecell{Temp. \\ (°C)} & \makecell{Density \\ (kg/m\textsuperscript{3})} & \makecell{Kinematic \\ Viscosity \\ (m\textsuperscript{2}/s)} & \makecell{Specific \\ Heat \\ (J/kg·K)} & \makecell{Thermal \\ Conductivity \\ (W/m·K)} & \makecell{Dynamic \\ Viscosity \\ (Pa·s)} \\
\hline
40 & 826.3 & 0.0000179 & 1980 & 0.14 & 0.01479077 \\
45 & 823.1 & 0.000015 & 2000  & 0.14 & 0.0123465 \\
50 & 820.0 & 0.0000127 & 2020 & 0.14 & 0.010414 \\
55 & 816.8 & 0.0000109 & 2040 & 0.14 & 0.00890312 \\
60 & 813.6 & 0.0000094 & 2060 & 0.14 & 0.00764784 \\
65 & 810.4 & 0.0000083 & 2080 & 0.13 & 0.00672632 \\
70 & 807.2 & 0.0000073 & 2090 & 0.13 & 0.00589256 \\
75 & 804.0 & 0.0000065 & 2110 & 0.13 & 0.005226 \\
80 & 800.8 & 0.0000058 & 2130 & 0.13 & 0.00464464 \\
85 & 797.5 & 0.0000052 & 2150 & 0.13 & 0.004147 \\
90 & 794.3 & 0.0000048 & 2170 & 0.13 & 0.00381264 \\
95 & 791.1 & 0.0000044 & 2190 & 0.13 & 0.00348084 \\
100 & 787.9 & 0.000004 & 2210 & 0.13 & 0.0031516 \\
105 & 784.6 & 0.0000037 & 2220 & 0.13 & 0.00290302 \\
110 & 781.4 & 0.0000034 & 2240 & 0.13 & 0.00265676 \\
115 & 778.1 & 0.0000032 & 2260 & 0.13 & 0.00248992 \\
120 & 774.9 & 0.000003 & 2280 & 0.13 & 0.0023247 \\
\hline
\end{tabularx}
\end{table}

\section{Methodology}

\subsection{Physics Models}
The study uses steady-state CFD models to capture the thermal and fluid dynamics of the different shaft designs and solves the governing equations for mass, momentum, and energy conservation with the finite volume method. We specifically use the transition shear stress transport (SST) model, which accounts for both laminar and turbulent flow regimes.

The used governing equations are as follows. The continuity equation is given by
\begin{equation}
\frac{\partial \rho}{\partial t} + \nabla \cdot (\rho \mathbf{v}) = 0,
\end{equation}
where $\rho$ is the fluid density, $t$ time, and $\mathbf{v}$  the velocity vector.
The momentum equation follows
\begin{equation}
\frac{\partial (\rho \mathbf{v})}{\partial t} + \nabla \cdot (\rho \mathbf{v} \mathbf{v}) = -\nabla p + \nabla \cdot (\tau) + \rho \mathbf{g},
\end{equation}
where $p$ is the pressure, $\tau$ the stress tensor, and $\mathbf{g}$ the gravitational acceleration.
The energy conservation reads
\begin{equation}
\frac{\partial (\rho E)}{\partial t} + \nabla \cdot \big(\mathbf{v} (\rho E + p)\big) = \nabla \cdot \big(k \nabla T + (\tau \cdot \mathbf{v})\big),
\end{equation}
where $E$ is the total energy, $k$ the thermal conductivity, and $T$ the temperature.

The finite volume method discretizes the equations and integrates the governing differential equations over each control volume in the mesh. The resulting algebraic equations are then solved iteratively to obtain the flow and temperature fields.

\subsection{Computational Setup}
We used a pressure-based solver and simulated the steady state with activated gravity just as the energy conservation equation to account for heat transfer.
The pressure-velocity coupling is handled using the coupled scheme. For spatial discretization, the gradient is calculated using the least squares cell-based method, while other parameters are discretized using the second order scheme. The pseudo time method uses a global time step to ensure stability and convergence of the solution.

\subsection{Mesh Definition and Convergence Analysis}
\begin{figure*}[htbp]
\centering
\includegraphics[width=\textwidth]{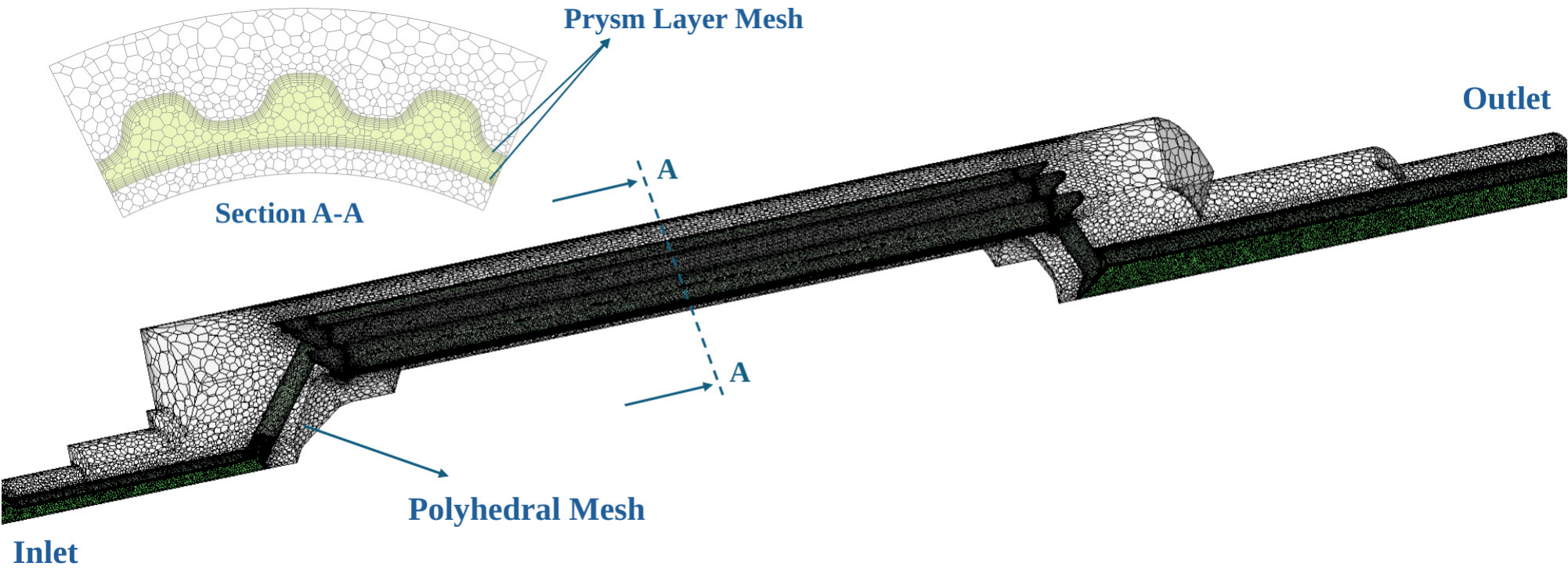}
\caption{Mesh configuration for Shaft Model 2 in ANSYS Fluent, illustrating the discretization used for simulation analysis.}
\label{fig:mesh1}
\end{figure*}

This study generated an unstructured polyhedral mesh with the ANSYS Fluent mesher to capture the complex geometry of the shafts. We tested multiple mesh densities to establish mesh independence, with the final configuration chosen to optimize the stability and reliability of simulation results. This approach allowed us to capture fine details at the boundaries where heat transfer interactions occur. Figure \ref{fig:mesh1} displays the mesh used for Shaft Model 2.

To leverage the circular symmetry of the shaft, we modeled only one seventh of the geometry, which reduced the computational requirements. This sector contained a total of 1,207,105 mesh elements. In this section, the transparent region represents the solid shaft, where only the energy equation was solved, and the green region corresponds to the fluid, where both momentum and energy equations were calculated.

\begin{figure}[htbp]
\centering
\includegraphics[width=0.45\textwidth]{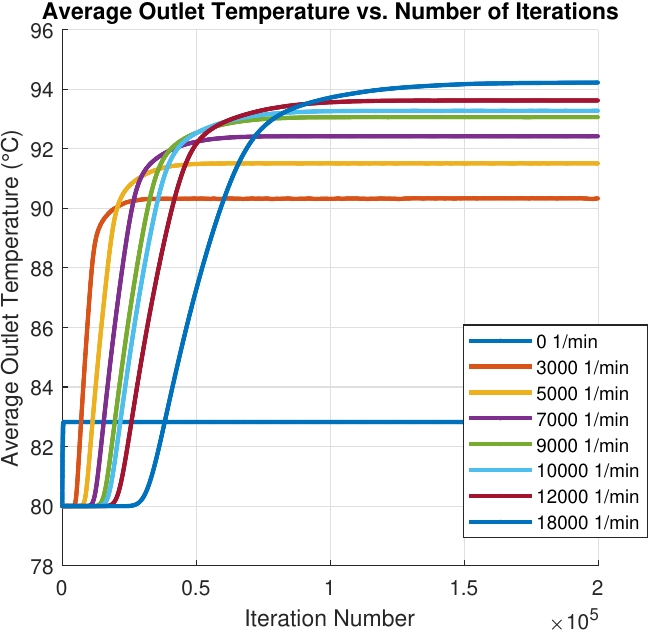}
\caption{Convergence of results for different rotational speeds.}
\label{fig:convergence}
\end{figure}

To ensure solution accuracy, we conducted a convergence analysis. We iterated simulations until the results stabilized and accurately reflected the fluid dynamics around the shaft. Figure \ref{fig:convergence} shows the convergence behavior across various rotational speeds. The stable convergence confirmed the reliability of predicted performance metrics.

\section{Results and Discussion}

\subsection{Impact of rotational speed}
We studied how changes in rotational speed affect the thermal and fluid dynamics of the shafts and evaluated outlet temperature, heat transfer rate, maximum pressure, and velocity profiles as key performance indicators. 

\begin{figure}[htbp]
\centering
\includegraphics[width=0.45\textwidth]{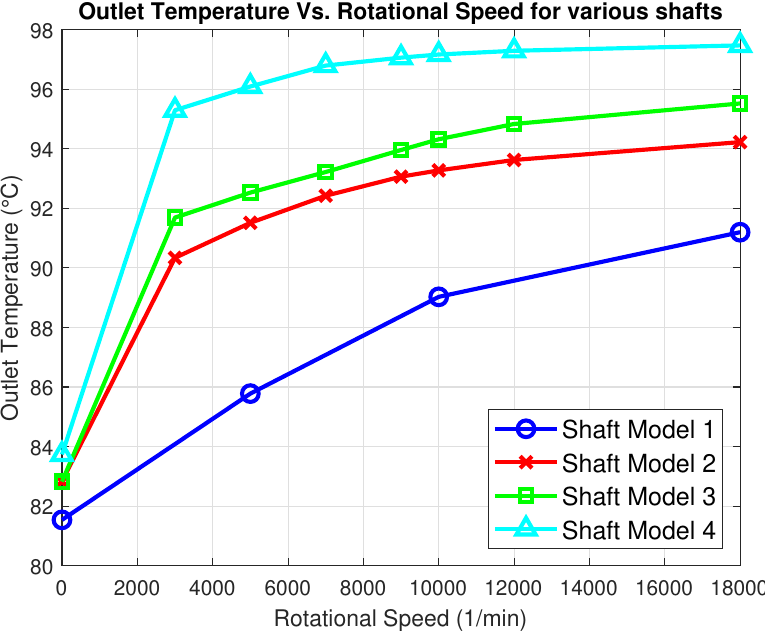}
\caption{Comparison of outlet temperature across all model shafts at various rotational speeds.}
\label{fig:outlet-temp-rpm}
\end{figure}

The outlet temperature of the shafts increases with the rotational speed (Fig.\ \ref{fig:outlet-temp-rpm}). This trend is consistent across different shaft designs and indicates that higher rotational speeds result in greater heat transfer to the fluid. Shaft Model 4 demonstrates the most significant temperature rise and suggests a better thermal absorption capability, presumably due to a larger surface area in contact with the fluid.

When the coolant passes through the shaft at a constant flow rate \(Q\), it is heated by the shaft’s internal wall. This process raises the coolant's temperature from \(T_{in}\) to \(T_{out}\). Per energy conservation, the rate of heat transfer \(P\) from the shaft to the coolant follows
\begin{equation}
    P = \rho \cdot Q \cdot c_p (T_{out} - T_{in}),
    \label{eq:heat_transfer_rate}
\end{equation}
where \(Q\) represents the volumetric flow rate (m\(^3\)/s), \(\rho\) is the coolant's density (kg/m\(^3\)), \(c_p\) is the specific heat capacity of the coolant (J/(kg·°C)), and \(T_\textrm{out}\) and \(T_\textrm{in}\) are respectively the outlet and inlet temperatures of the coolant (°C).

\begin{figure}[htbp]
\centering
\includegraphics[width=0.45\textwidth]{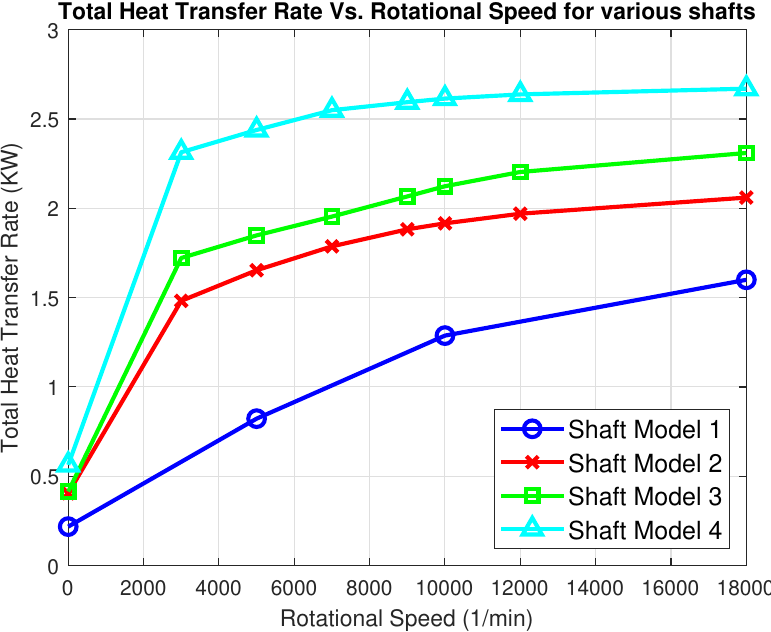}
\caption{Comparison of total heat transfer rate across all model shafts at various rotational speeds.}
\label{fig:heat-transfer-rpm}
\end{figure}

As the rotational speed increases, so does the total heat transfer rate (Fig.\ \ref{fig:heat-transfer-rpm}). Across all designs, Shaft Model 1 has the lowest cooling performance while Shaft Model 4 performs the best. This comparison reflects that Shaft Model 4's surface area for heat transfer is more than doubled compared to Shaft Model 1 despite equal overall size of the shafts (Table \ref{tab:rotor_shaft_dimensions}). Shafts Models 2 and 3 also saw a surface area increase, but only by about 15\%. 

\begin{figure}[htbp]
\centering
\subfloat[First Model Shaft]{\includegraphics[width=0.24\textwidth]{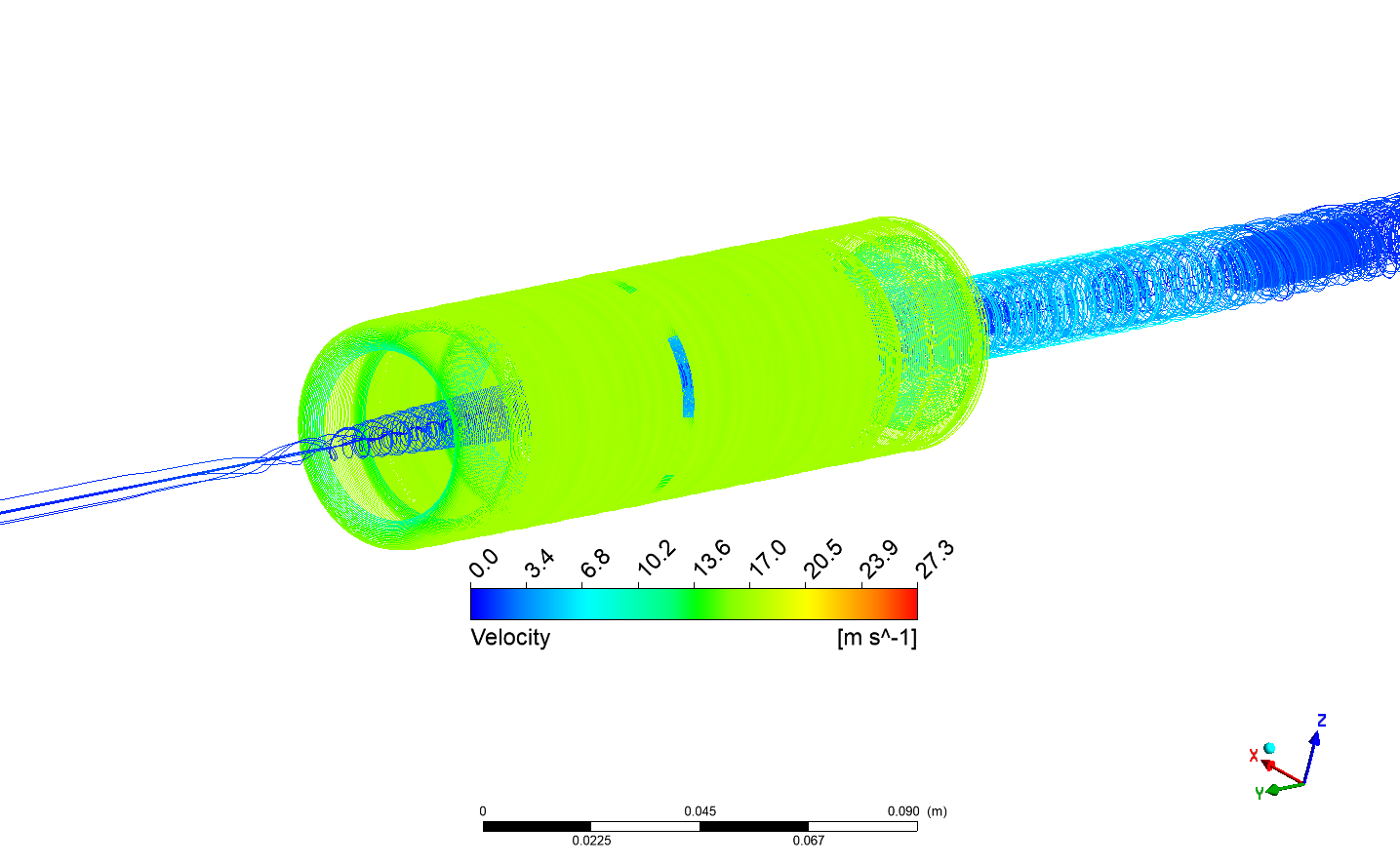}\label{fig:stream-model1-rpm-10000-comp}}
\hfill
\subfloat[Second Model Shaft]{\includegraphics[width=0.24\textwidth]{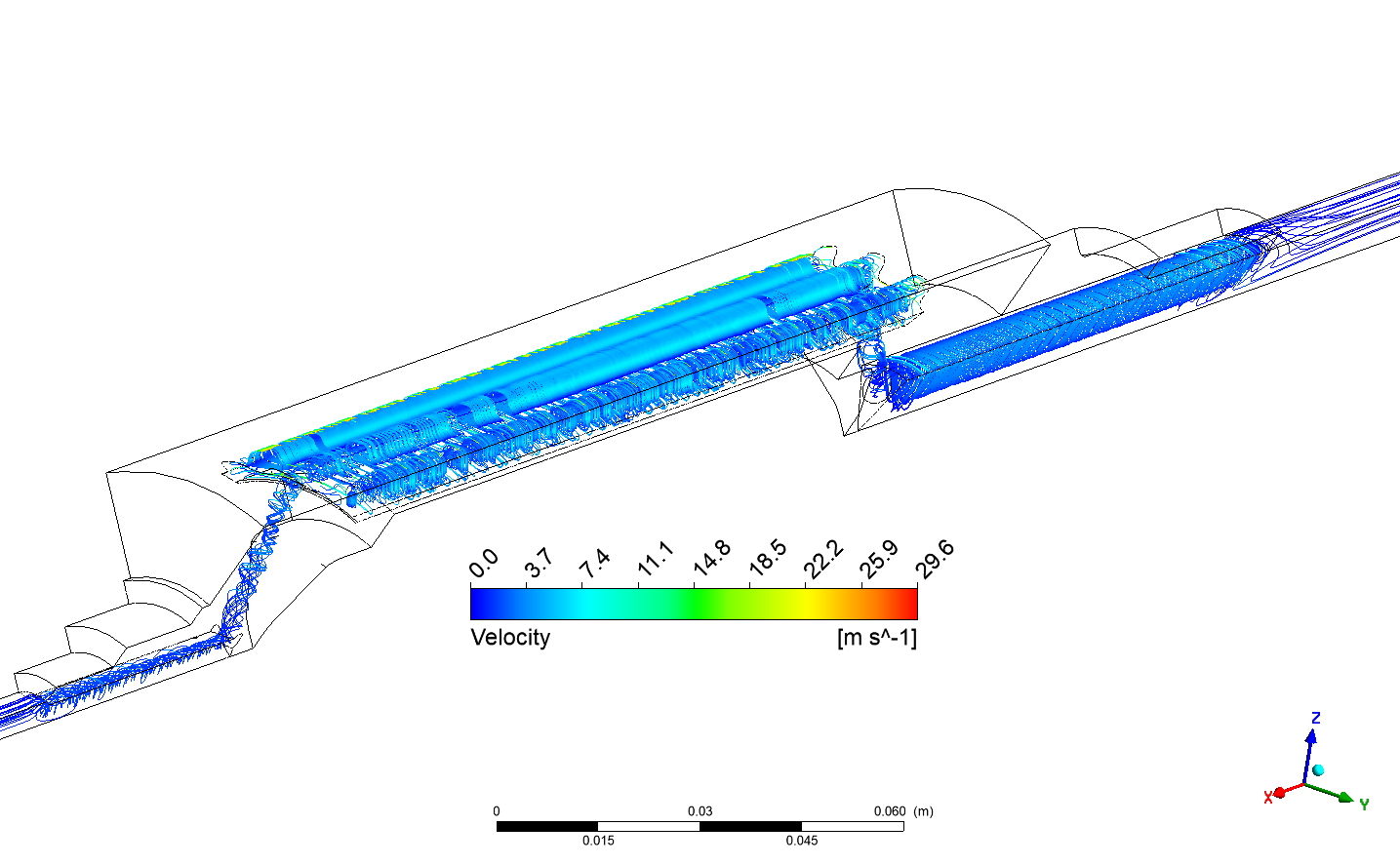}\label{fig:stream-model2-rpm-10000-comp}}
\hfill
\subfloat[Third Model Shaft]{\includegraphics[width=0.24\textwidth]{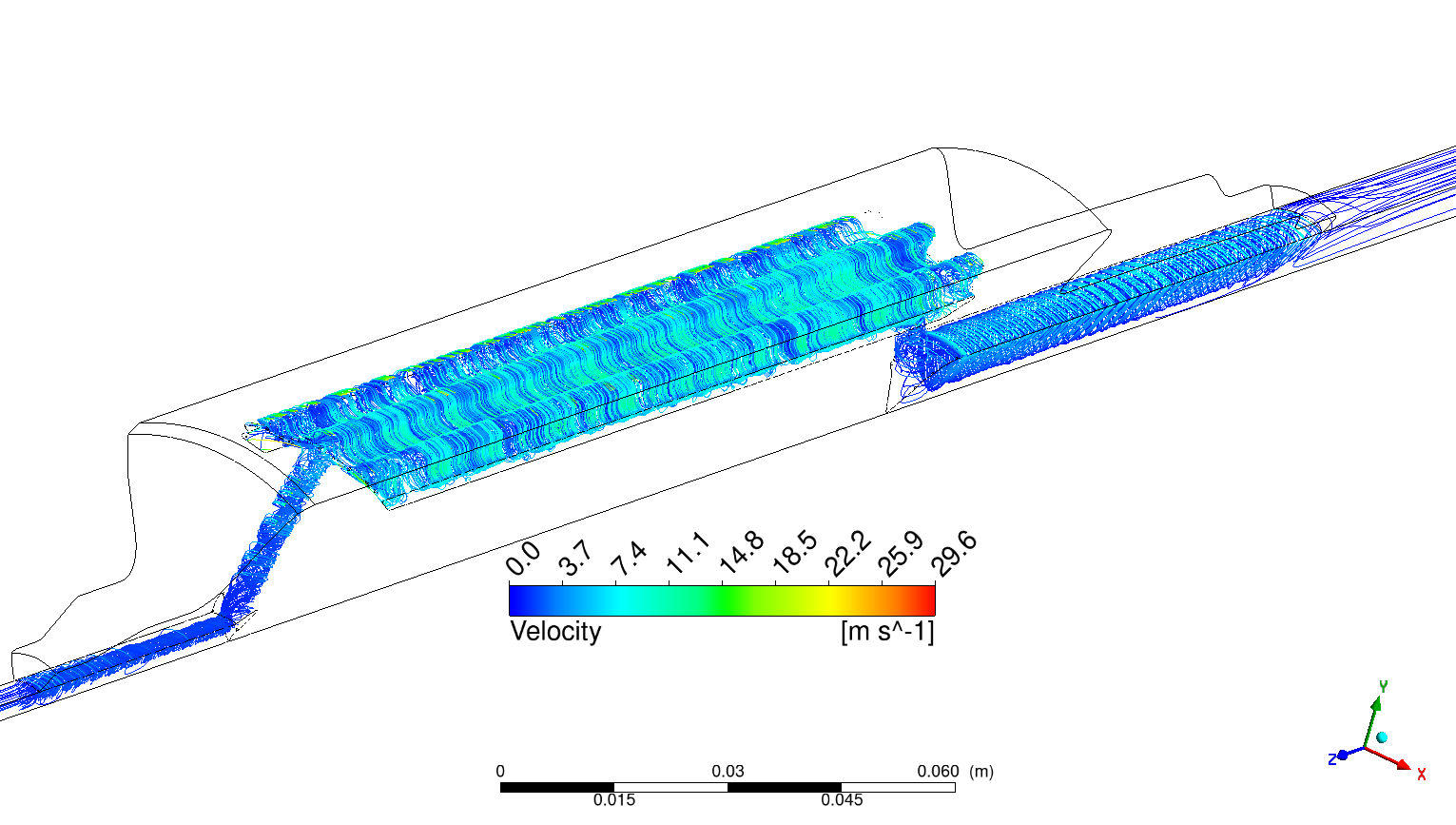}\label{fig:stream-model3-rpm-10000-comp}}
\hfill
\subfloat[Fourth Model Shaft]{\includegraphics[width=0.24\textwidth]{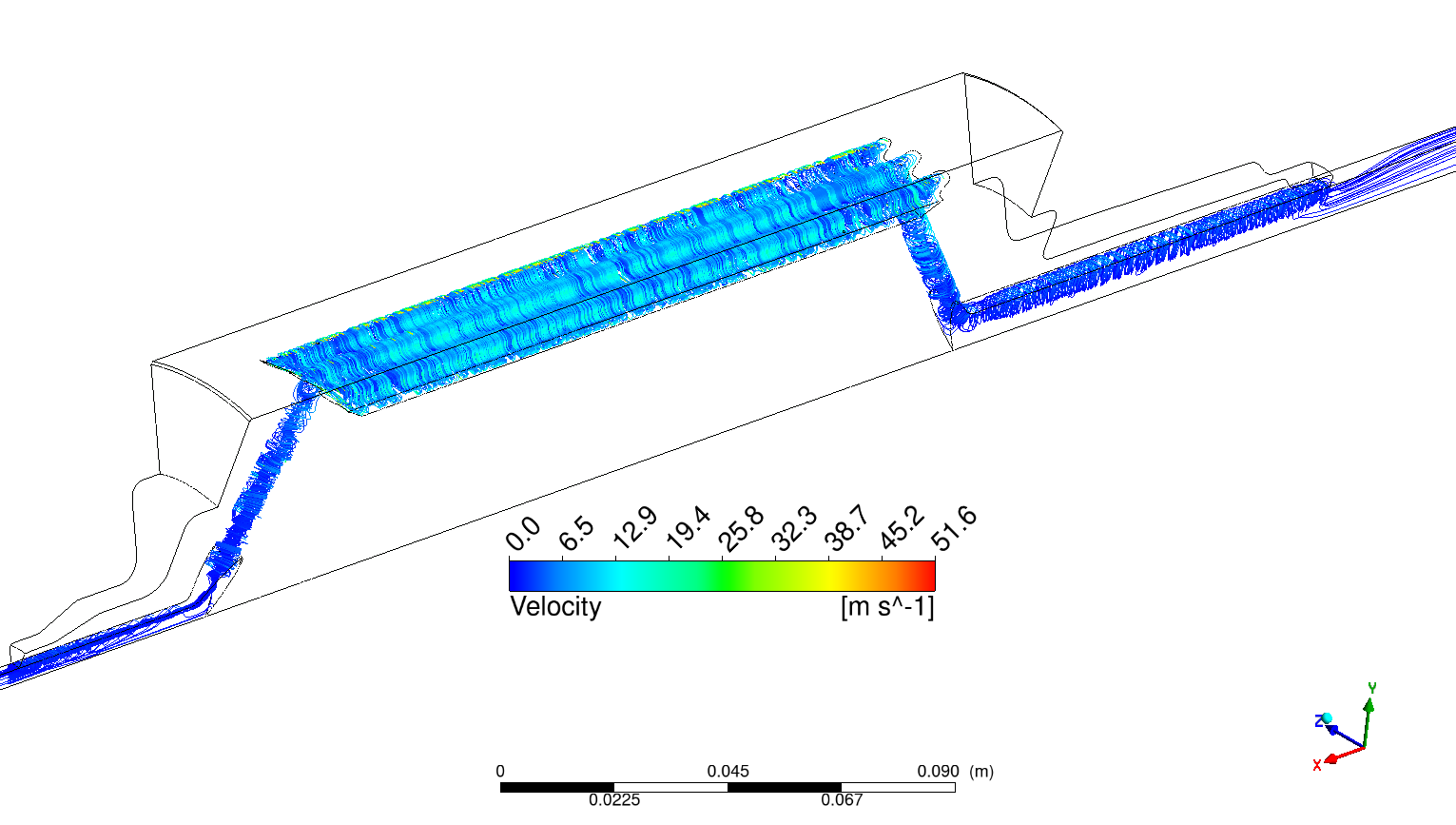}\label{fig:stream-model4-rpm-10000-comp}}
\caption{Isometric view of streamlines of various shaft models at 10,000 1/min.}
\label{fig:stream-rpm-10000-comps}
\end{figure}

\begin{figure}[htbp]
\centering
\subfloat[First Model Shaft]{\includegraphics[trim={7cm 6cm 7cm 9cm},clip,width=0.24\textwidth]{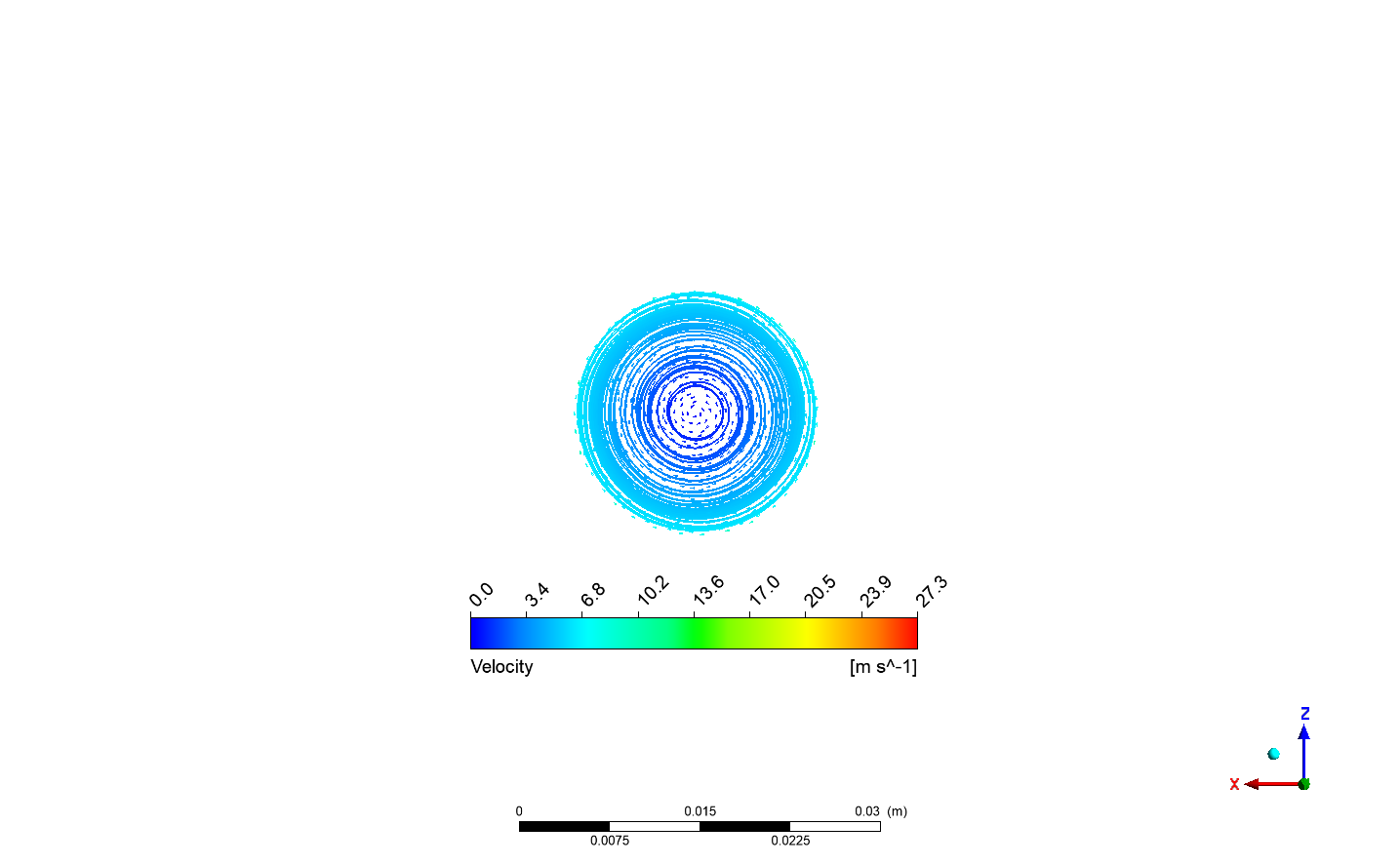}\label{fig:stream-cross-model1-rpm-10000-comp}}
\hfill
\subfloat[Second Model Shaft]{\includegraphics[trim={3cm 6cm 4cm 9cm},clip,width=0.24\textwidth]{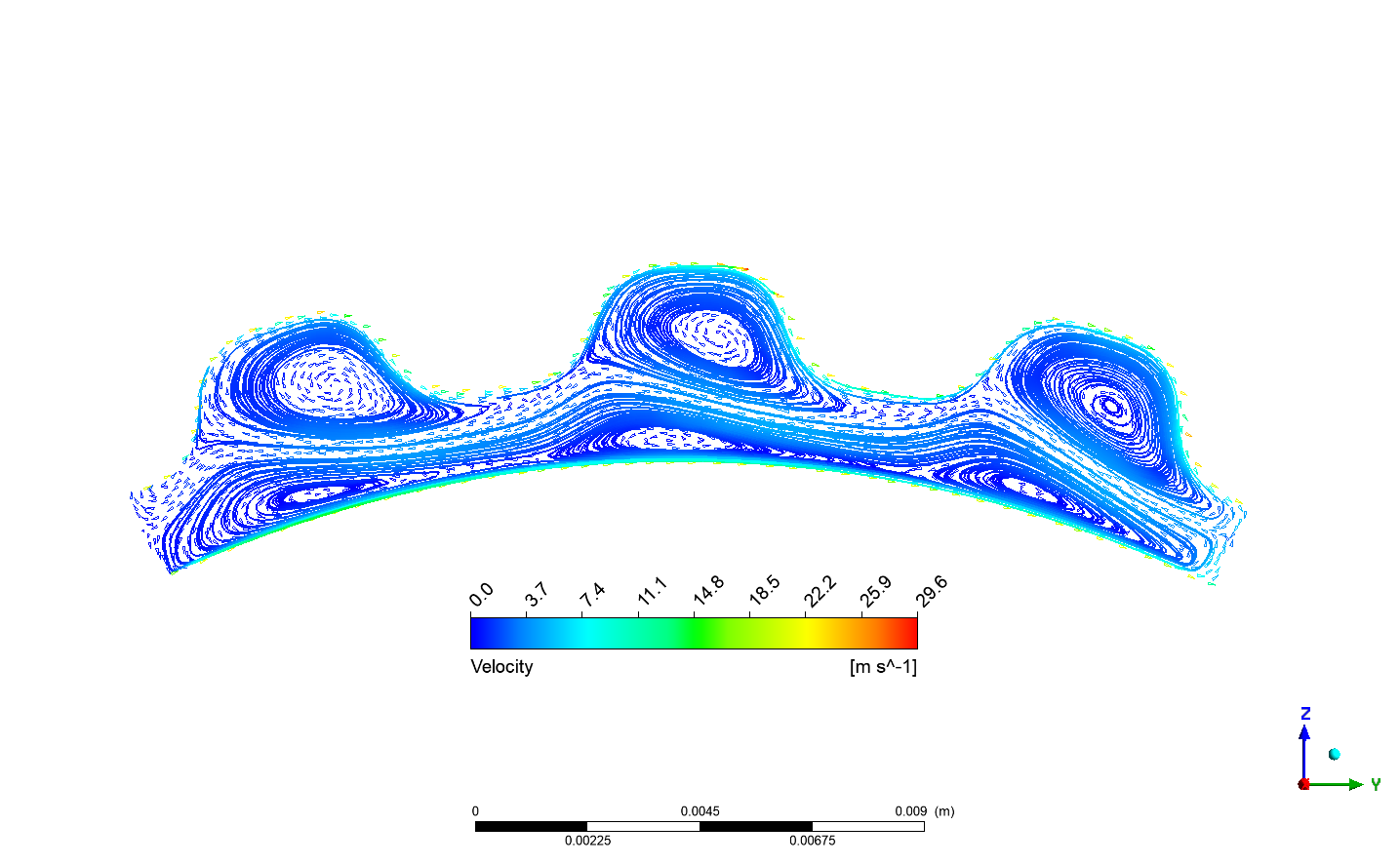}\label{fig:stream-cross-model2-rpm-10000-comp}}
\hfill
\subfloat[Third Model Shaft]{\includegraphics[trim={3cm 6cm 4cm 3cm},clip,width=0.24\textwidth]{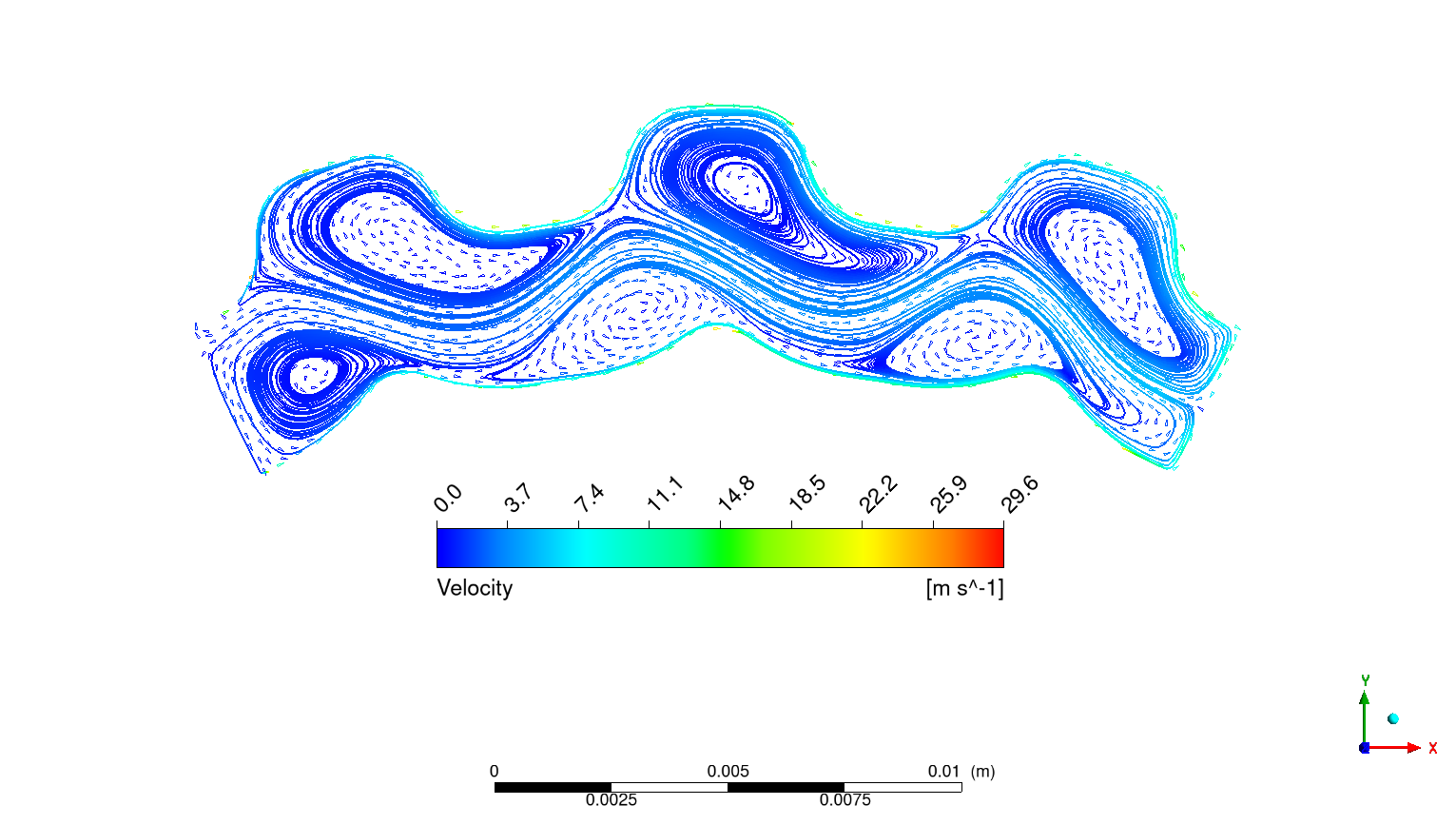}\label{fig:stream-cross-model3-rpm-10000-comp}}
\hfill
\subfloat[Fourth Model Shaft]{\includegraphics[trim={7cm 6cm 7cm 6cm},clip,width=0.24\textwidth]{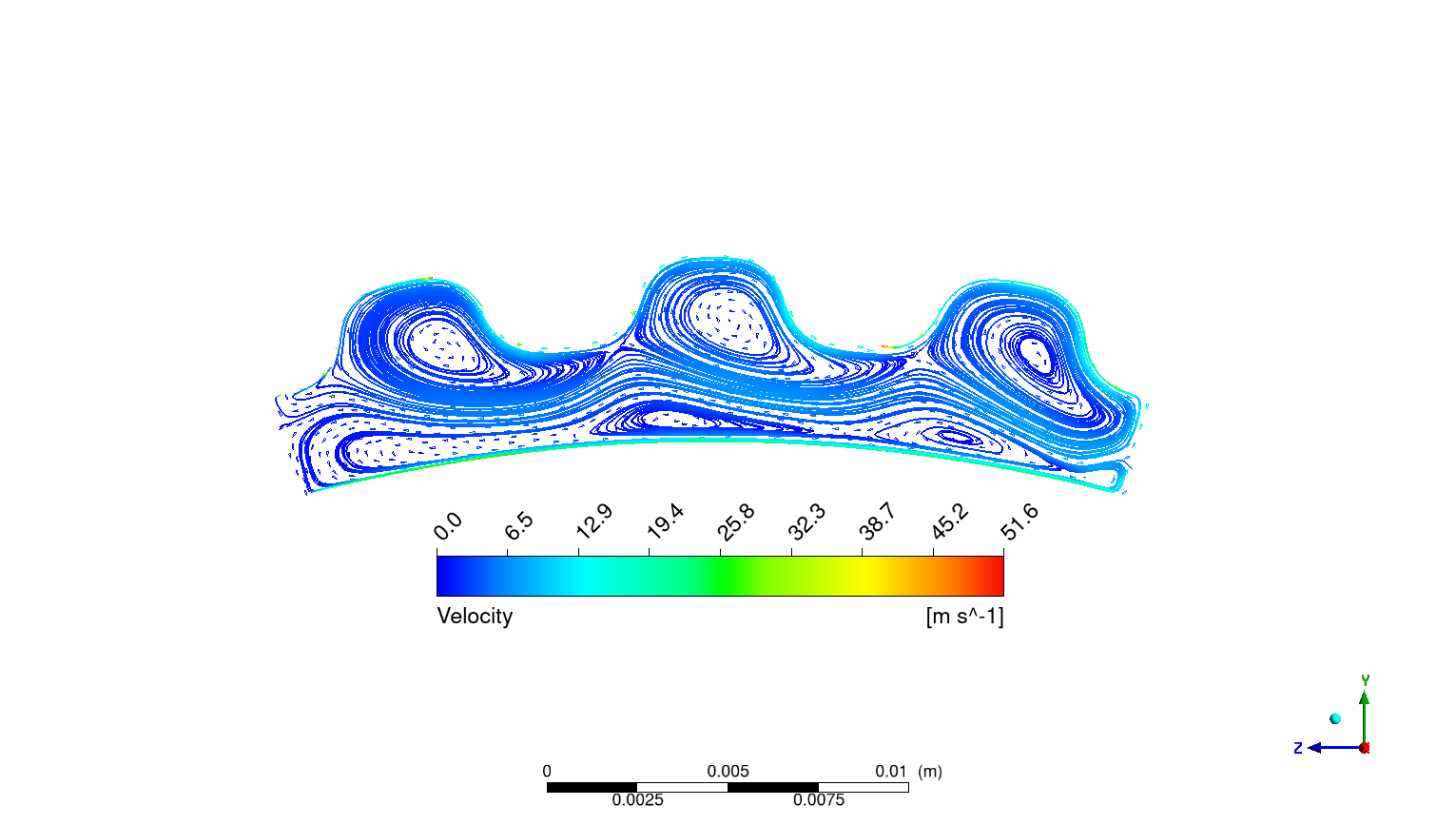}\label{fig:stream-cross-model4-rpm-10000-comp}}
\caption{Cross-sectional view of streamlines of various shaft models at 10,000 1/min.}
\label{fig:stream-cross-rpm-10000-comps}
\end{figure}

In addition to mere surface area, turbulent flow, which causes mixing of the hot fluid near the wall with the cooler fluid moving through the center, also enhances cooling. This effect is highlighted in Figures \ref{fig:stream-rpm-10000-comps} and \ref{fig:stream-cross-rpm-10000-comps}, which show the flow patterns around all four shafts at 10,000 1/min.

\begin{figure}[htbp]
\centering
\includegraphics[width=0.45\textwidth]{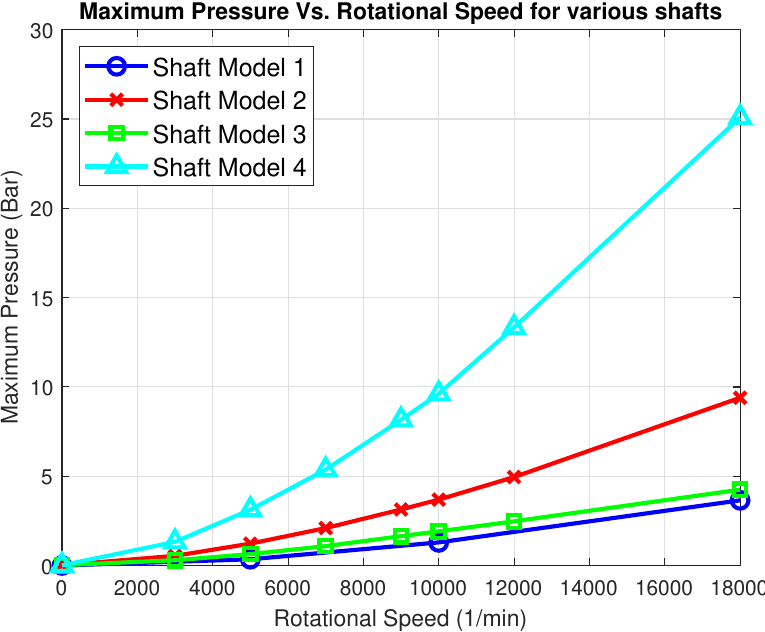}
\caption{Comparison of maximum pressure across all model shafts at various rotational speeds.}
\label{fig:max-press-rpm}
\end{figure}

Figure \ref{fig:max-press-rpm} illustrates that the  pressure within the shafts, which the pump has to overcome, also increases with the rotational speed. The pressure rise is particularly noticeable in Shaft Model 4 at higher speeds.

\begin{figure}[htbp]
\centering
\includegraphics[width=0.45\textwidth]{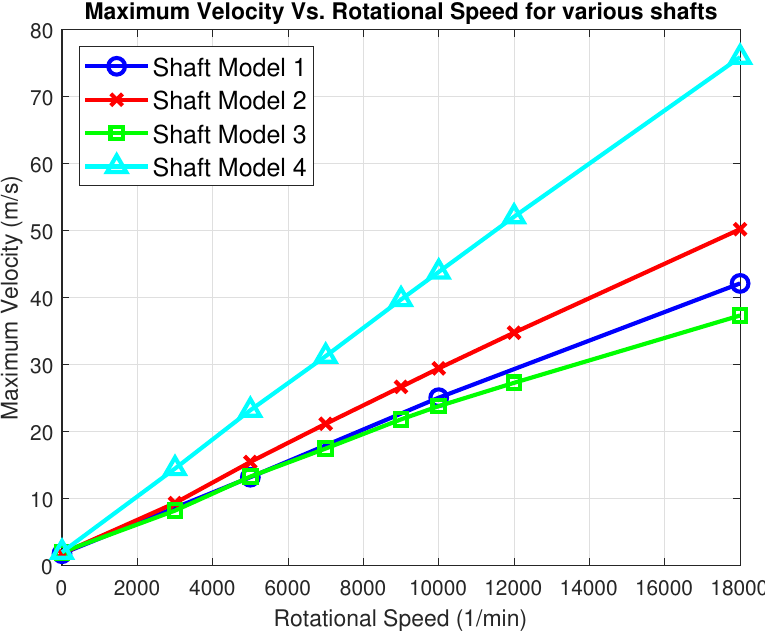}
\caption{Comparison of maximum coolant velocity across all model shafts at various 1/min.}
\label{fig:max-vel-rpm}
\end{figure}

Figure\ref{fig:max-vel-rpm} illustrates the relationship between the maximum coolant velocity in the shaft channels and the shaft’s rotational speed.
The fluid flow within a rotating cylinder is substantially determined by the associated speed-dependent rotational forces, specifically the centrifugal and Coriolis forces. The centrifugal force acts outwards and affects the radial distribution of pressure and velocity within the fluid.
For an infinitesimal fluid element with volume \(\textrm{d}V\) and density \(\rho\), the centrifugal force per unit volume follows
\begin{equation}
    \textrm{d}F_{c} = \rho\, \textrm{d}V\, \omega^{2}\, r,
\end{equation}
where \(r\) is the radial distance from the rotational axis and \(\omega\) is the angular velocity. In addition to the centrifugal force, the Coriolis force acts tangentially and radially to the rotation. This force emerges with the cylinder's rotation and acts perpendicularly to the direction of fluid flow and the rotation axis. It contributes to the secondary flow patterns within the rotating cylinder per
\begin{equation}
    \textrm{d}F_{\text{cor}} = 2 \rho\, \textrm{d}V\, \left(\omega \times \mathbf{v}\right)
\end{equation}

As a consequence, the maximum coolant velocity inside the shaft increases linearly with the shaft rotational speed $\omega$ per \(V = r \cdot \omega\). Conversely, the maximum pressure exhibits a parabolic trend as does the centrifugal force.

As the rotational speed increases, we observe higher outlet temperatures,  heat transfer rates, and velocities but also increased pressure for the pump.

\subsection{Impact of Inlet Volumetric Flow Rate}
\begin{figure}[htbp]
\centering
\includegraphics[width=0.45\textwidth]{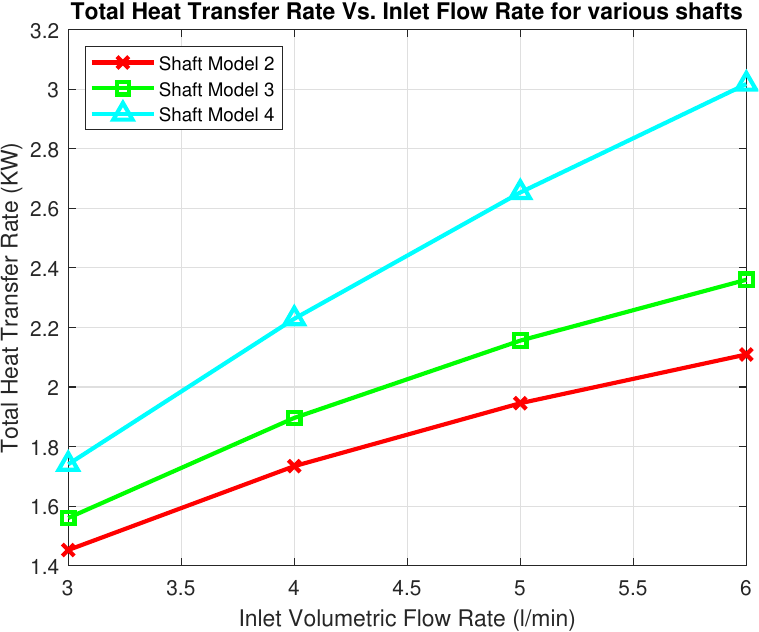}
\caption{Comparison of total heat transfer rate across all model shafts at various inlet volumetric flow rate.}
\label{fig:heat-inflowrate}
\end{figure}

We studied the impact of the inlet flow rate on the fluid dynamics and the thermal performance at a constant rotational speed of 10,000 1/min. Although a higher flow rate reduces the residence time of the coolant in the shaft, the heat-transfer rate increases with the flow rate (Fig.\ \ref{fig:heat-inflowrate}). The increased flow carries a larger mass of fluid that extracts more heat from the shaft. Shaft Model 4 exhibits distinctly higher heat-transfer rate compared to the others due to its larger outer diameter and increased heat transfer surface area.


\begin{figure}[htbp]
\centering
\includegraphics[width=0.45\textwidth]{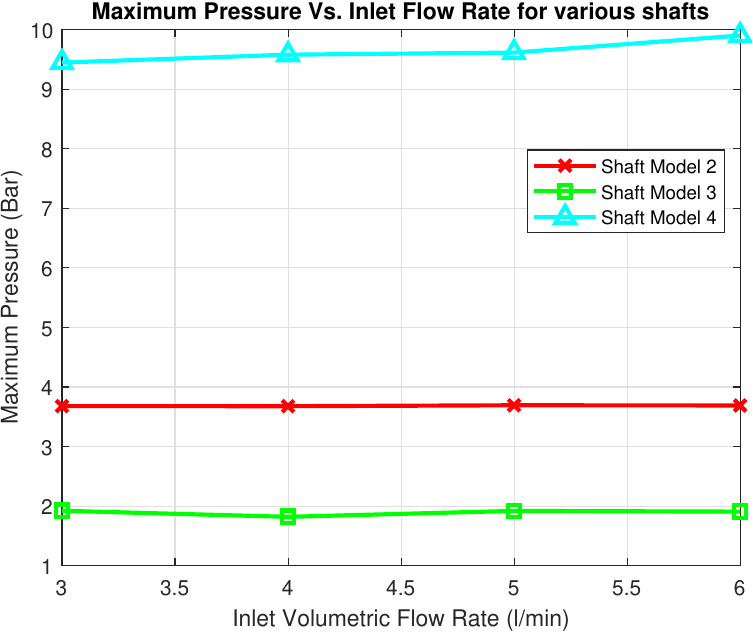}
\caption{Comparison of maximum pressure across all model shafts at various inlet volumetric flow rate.}
\label{fig:max-press-inflowrate}
\end{figure}

The pressure remains relatively constant over the inlet flow rate (Fig.\ \ref{fig:max-press-inflowrate}). Only for Shaft Model 4, the pressure increases moderately with the flow rate. Notably, Shaft Model 4 exhibits two to five times higher maximum pressure compared to the others due to its larger outer diameter but similar cross section of the ducts as that of Shaft Model 2.

Although Shaft 4 has a substantially larger outer diameter (about 98 mm) compared to Shaft 2 (56.6 mm), this difference alone does not explain the significant increase in pressure. Importantly, the per-channel cross-sectional area in Shaft 4 is comparable to Shaft 2’s, meaning that the nominal flow area per channel is not the critical factor. Instead, the primary differences stem from the number of channels (36 in Shaft Model 4 vs. 21 in Shaft Model 2) and their distribution around the circumference.

By increasing the number of channels, Shaft Model 4 inevitably increases the total wetted perimeter through which the coolant interacts with the channel surfaces. A higher wetted perimeter leads to greater frictional losses, particularly because the fluid experiences more frequent interruptions and boundary interactions as it navigates through the densely packed channels.

Furthermore, the presence of more channels implies a longer cumulative flow path, particularly when considering the rotational dynamics at play. While the total cross-sectional area might be similar, the effective path length through which the coolant travels is longer, which results in higher pressure loss due to increased friction.

Streamline plots reveal localized velocity spikes of 40–50\,m/s in Shaft Model 4, which result from the numerous tooth boundaries forcing rapid directional changes (Figs.\ \ref{fig:stream-rpm-10000-comps} and \ref{fig:stream-cross-rpm-10000-comps}). This excessive turbulence and frequent acceleration around the channels contribute to additional dynamic pressure losses, particularly at higher rotational speeds.

In contrast, Shaft Model 3 achieves significantly lower pressures ($\approx 2$\,bar at 10,000\,rpm) despite demonstrating higher heat transfer performance. This is attributable to its wavy, more open channel design, which allows efficient mixing while minimizing frictional losses. Shaft Model 2, with its 21 channels, shows moderately higher pressure levels than Shaft Model 3 but remains far below those of Shaft Model 4.

\begin{figure}[htbp]
\centering
\includegraphics[width=0.45\textwidth]{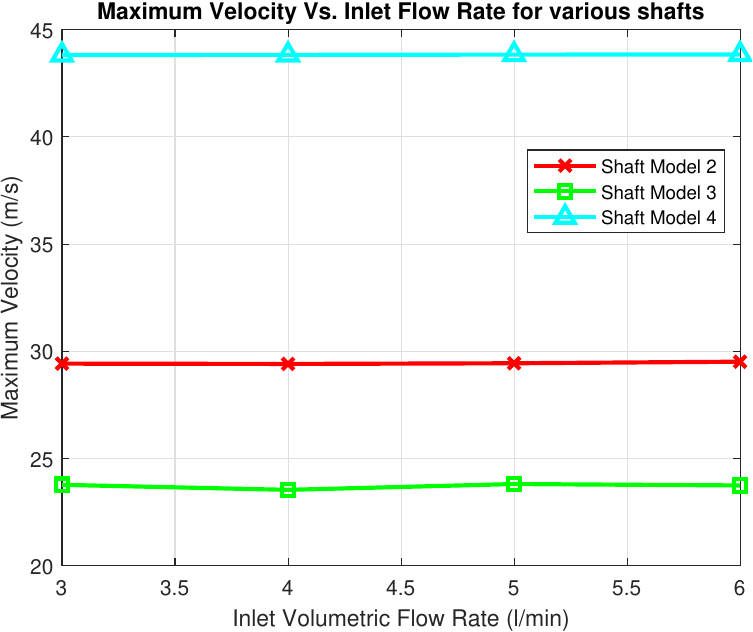}
\caption{Comparison of maximum coolant velocity across all model shafts at various inlet volumetric flow rate.}
\label{fig:max-vel-inflowrate}
\end{figure}

Figure~\ref{fig:max-vel-inflowrate} confirms that, although the coolant velocity rises slightly with increased inlet flow, the geometry of each shaft overwhelmingly dictates its velocity profile. Shaft Model 4 attains the highest local velocities not only due to its larger diameter, but also due to the high channel count and increased wetted perimeter, which enhance surface friction and require greater pumping pressure.

\subsection{Impact of Inlet Temperature}

\begin{figure}[htbp]
\centering
\subfloat[T\textsubscript{in}=50\textdegree{}C]{\includegraphics[trim={4cm 6cm 4cm 6cm},clip, width=0.24\textwidth]{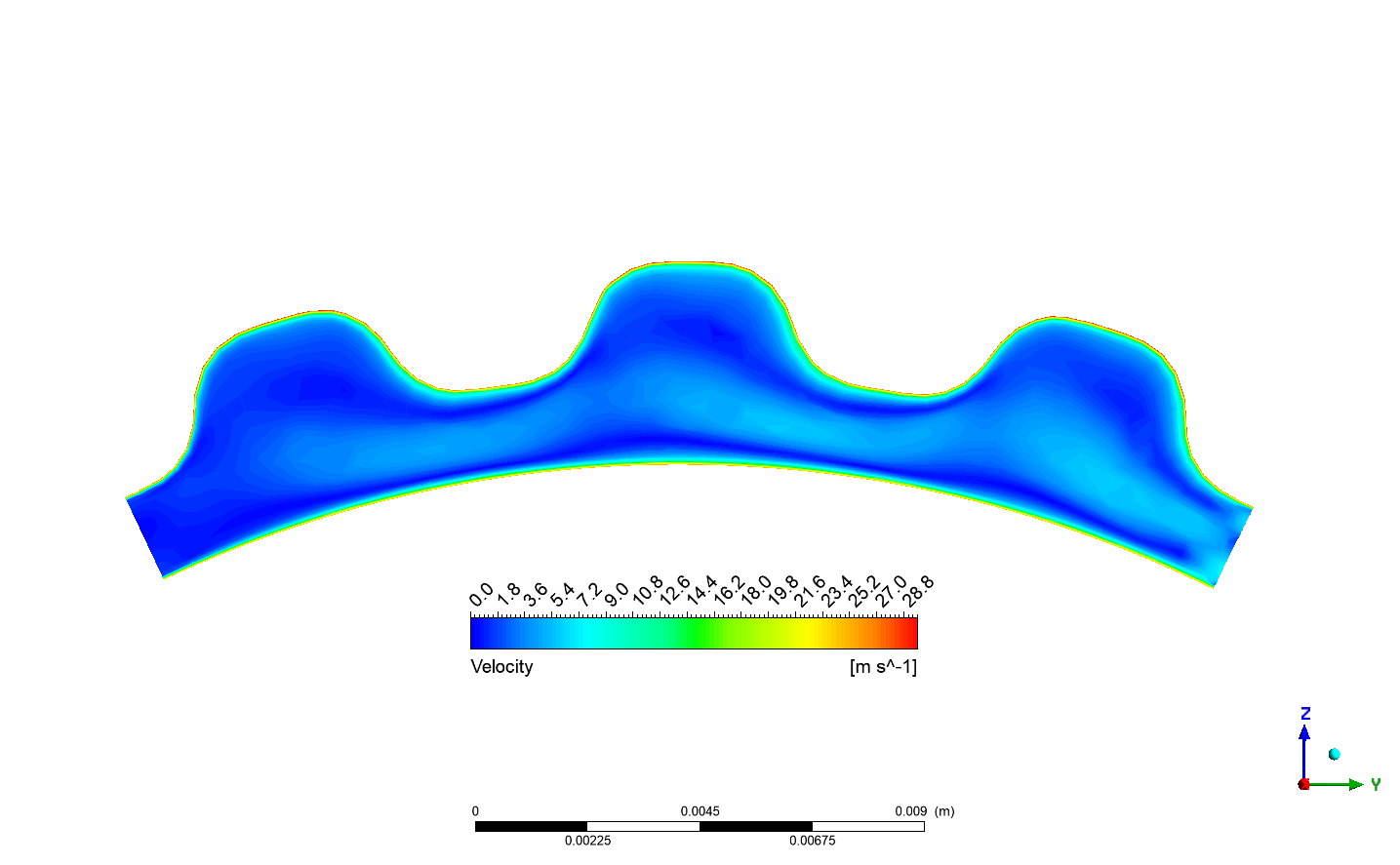}\label{fig:velocity-cross-model2-rpm-10000-intemp50}}
\hfill
\subfloat[T\textsubscript{in}=60\textdegree{}C]{\includegraphics[trim={3cm 6cm 3cm 6cm},clip, width=0.24\textwidth]{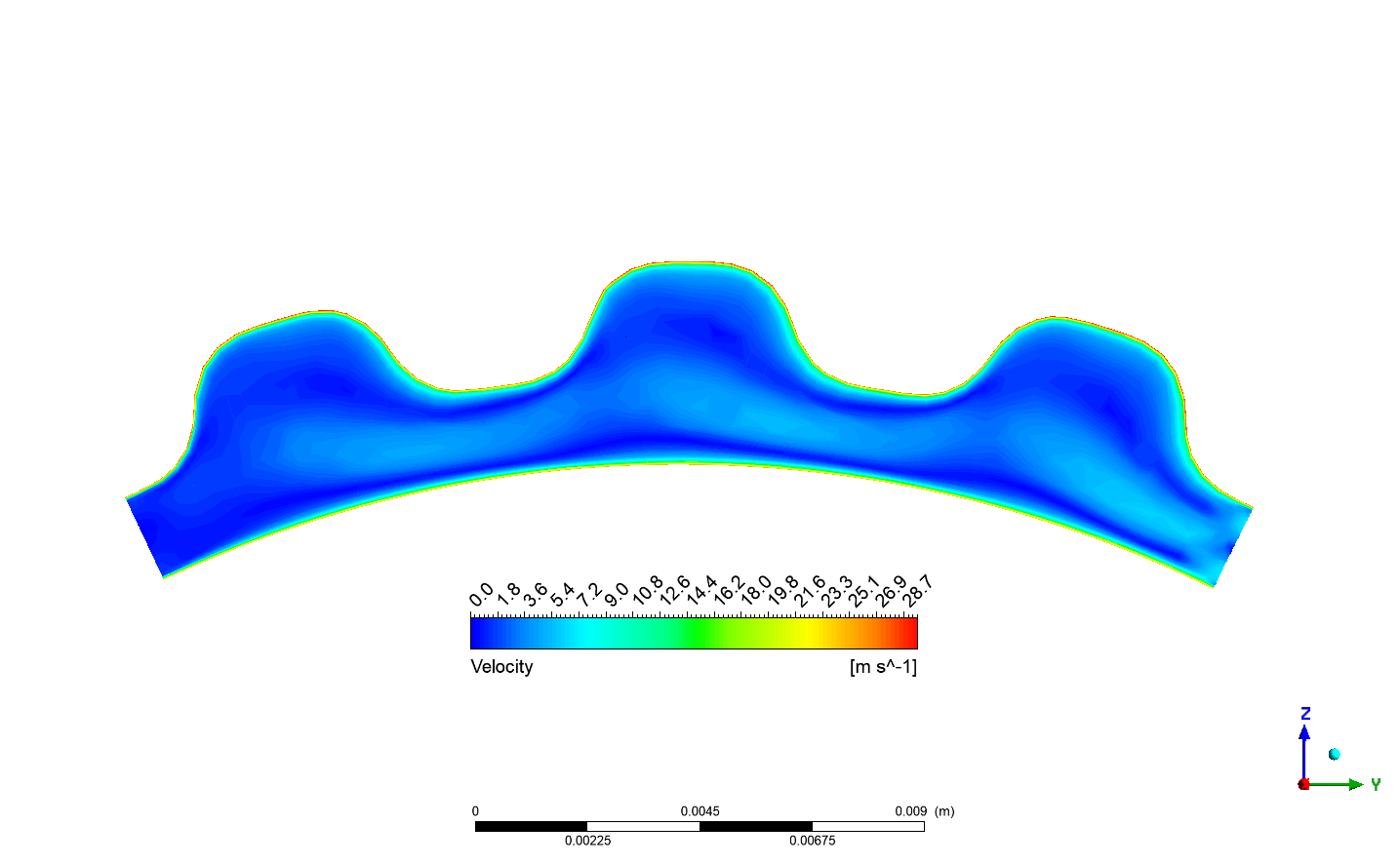}\label{fig:velocity-cross-model2-rpm-10000-intemp60}}
\hfill
\subfloat[T\textsubscript{in}=70\textdegree{}C]{\includegraphics[trim={3cm 6cm 3cm 6cm},clip, width=0.24\textwidth]{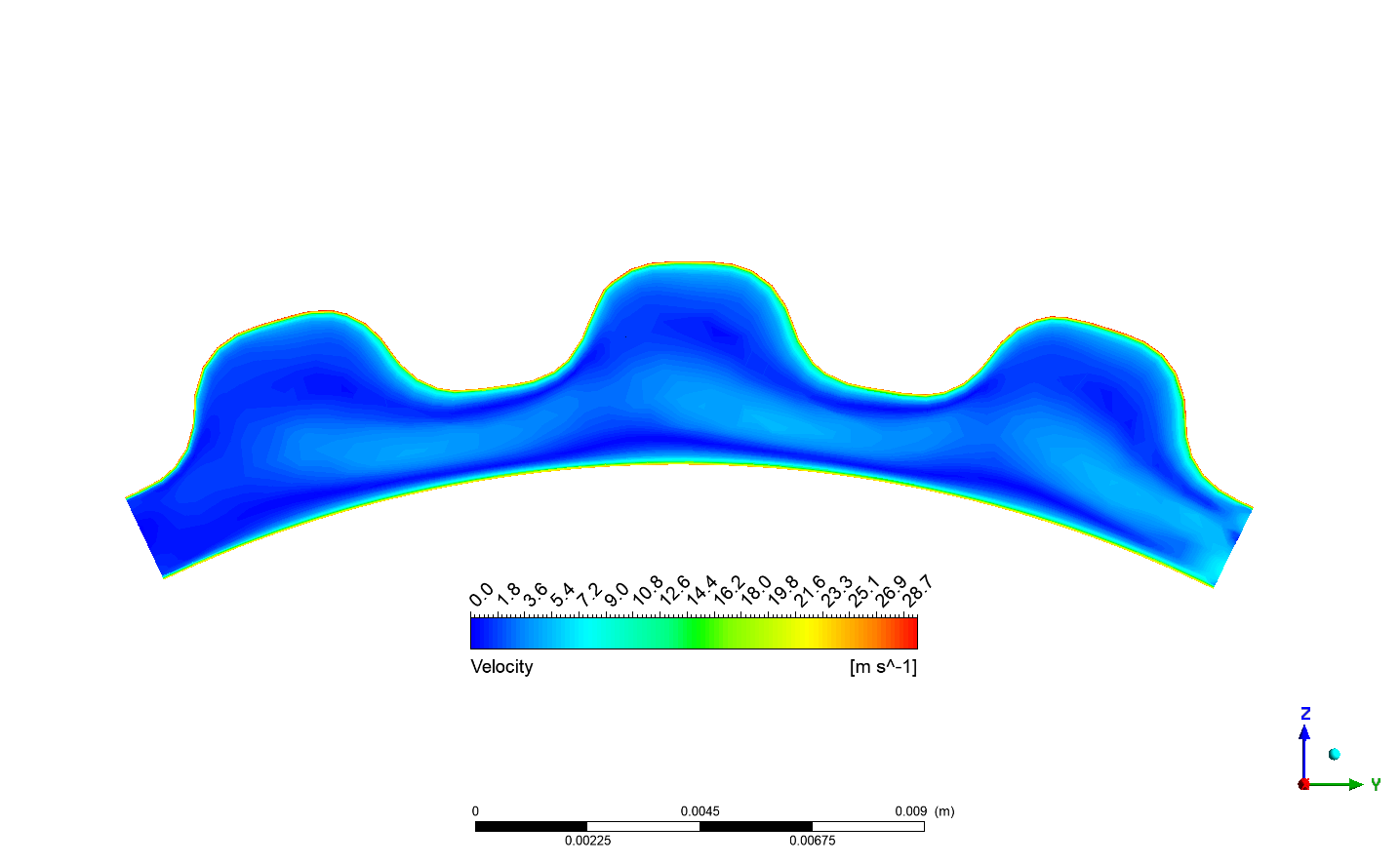}\label{fig:velocity-cross-model2-rpm-10000-intemp70}}
\hfill
\subfloat[T\textsubscript{in}=80\textdegree{}C]{\includegraphics[trim={3cm 6cm 3cm 6cm},clip, width=0.24\textwidth]{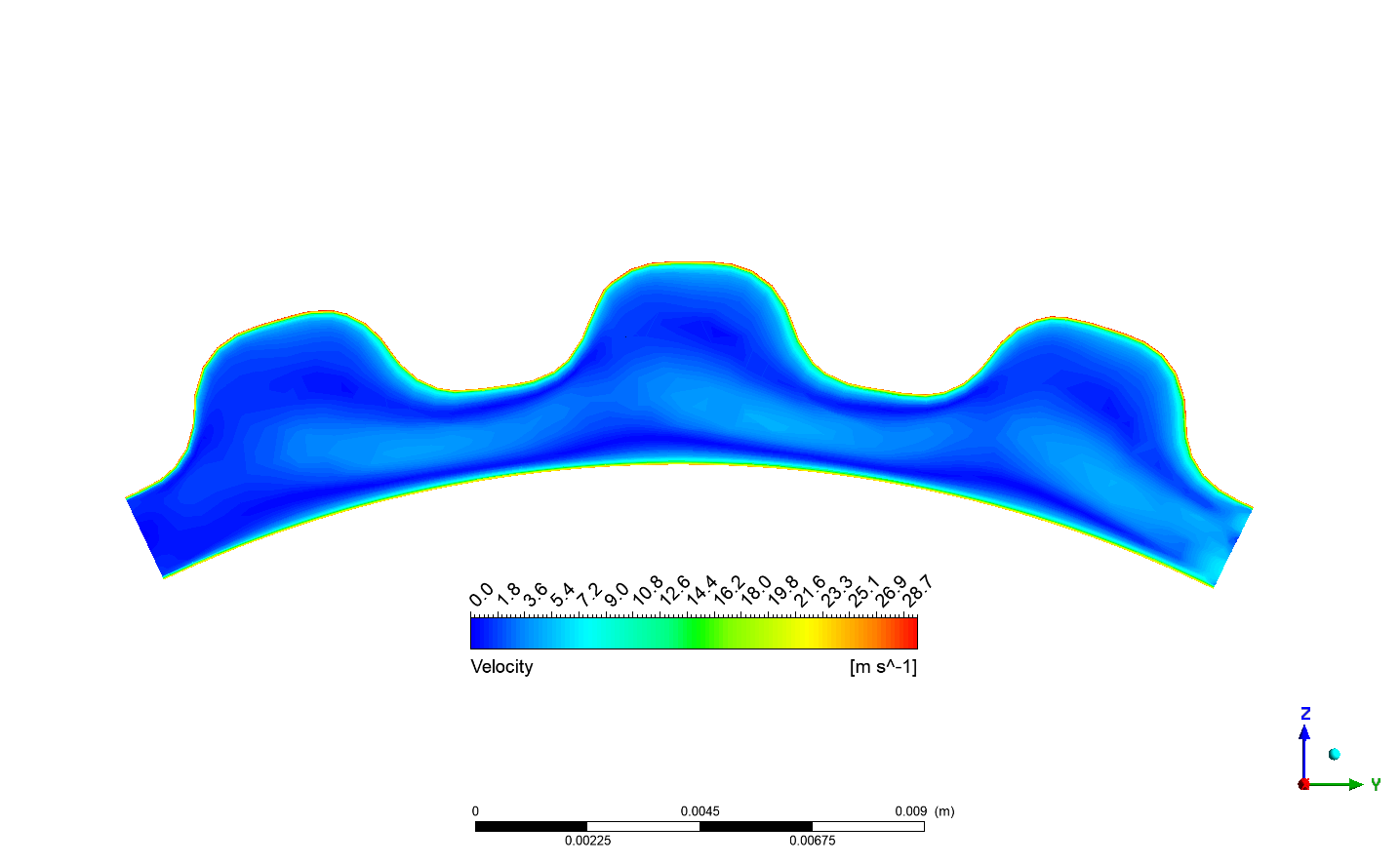}\label{fig:velocity-cross-model2-rpm-10000-intemp80}}
\caption{Cross-sectional view of velocity contours for Shaft Model 2 at 10,000 1/min across different inlet temperatures.}
\label{fig:velocity-cross-model2-intemp-comps-section3}
\end{figure}

Increasing inlet temperature reduces the coolant's viscosity, which can alter the flow and pressure characteristics.
For Shaft Model 2, velocity contours remain nearly constant as inlet temperatures varies (Fig.\ \ref{fig:velocity-cross-model2-intemp-comps-section3}). The flow pattern through its toothed outer wall and smooth inner tube does not change significantly. The geometry enforces a uniform flow path. Lower viscosity mainly reduces frictional losses without substantially redistributing the velocity field.

\begin{figure}[htbp]
\centering
\subfloat[T\textsubscript{in}=50\textdegree{}C]{\includegraphics[trim={4cm 6cm 4cm 3cm},clip, width=0.24\textwidth]{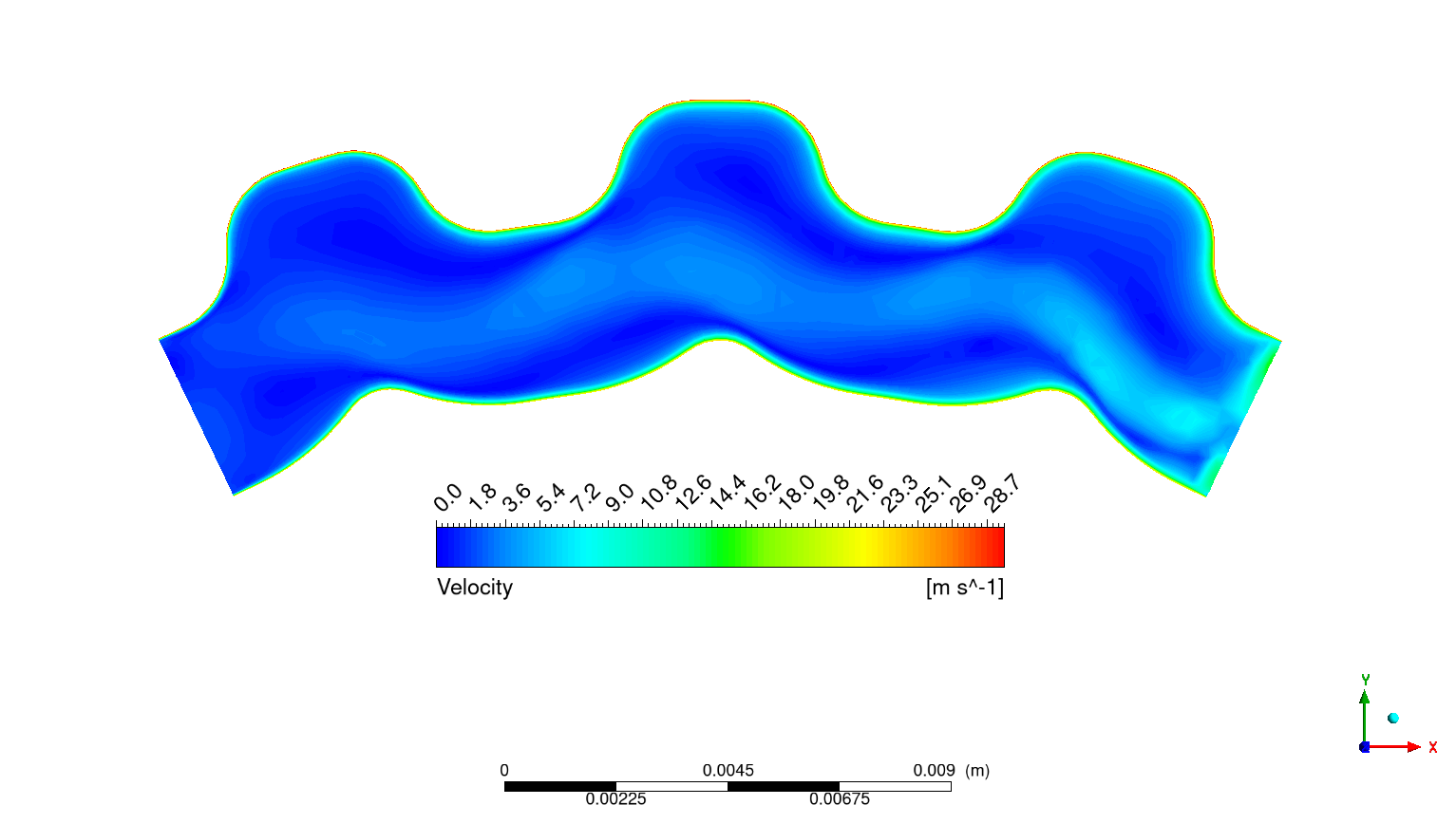}\label{fig:velocity-cross-model3-rpm-10000-intemp50}}
\hfill
\subfloat[T\textsubscript{in}=60\textdegree{}C]{\includegraphics[trim={6cm 6cm 6cm 3cm},clip, width=0.24\textwidth]{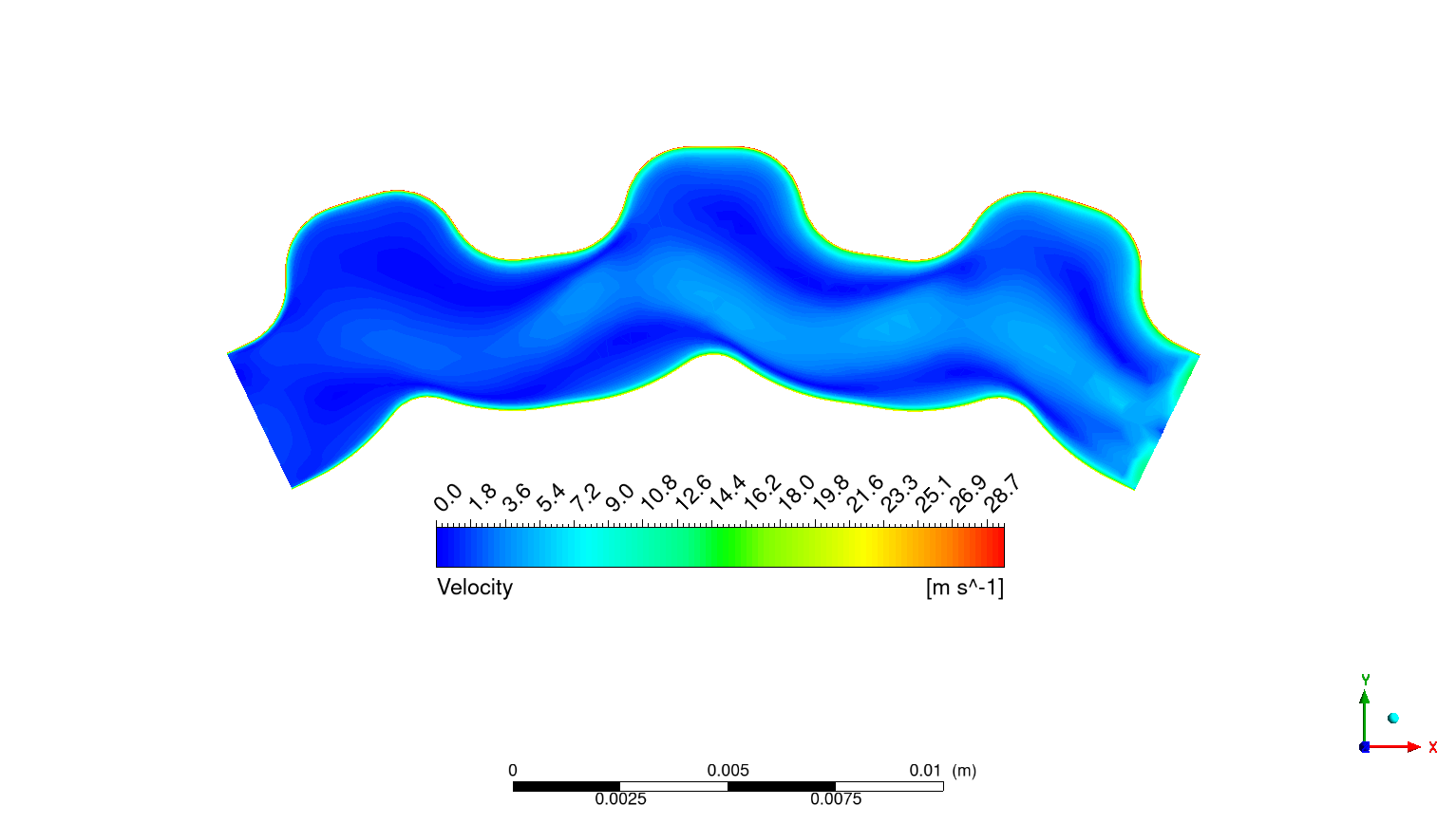}\label{fig:velocity-cross-model3-rpm-10000-intemp60}}
\hfill
\subfloat[T\textsubscript{in}=70\textdegree{}C]{\includegraphics[trim={6cm 6cm 6cm 3cm},clip, width=0.24\textwidth]{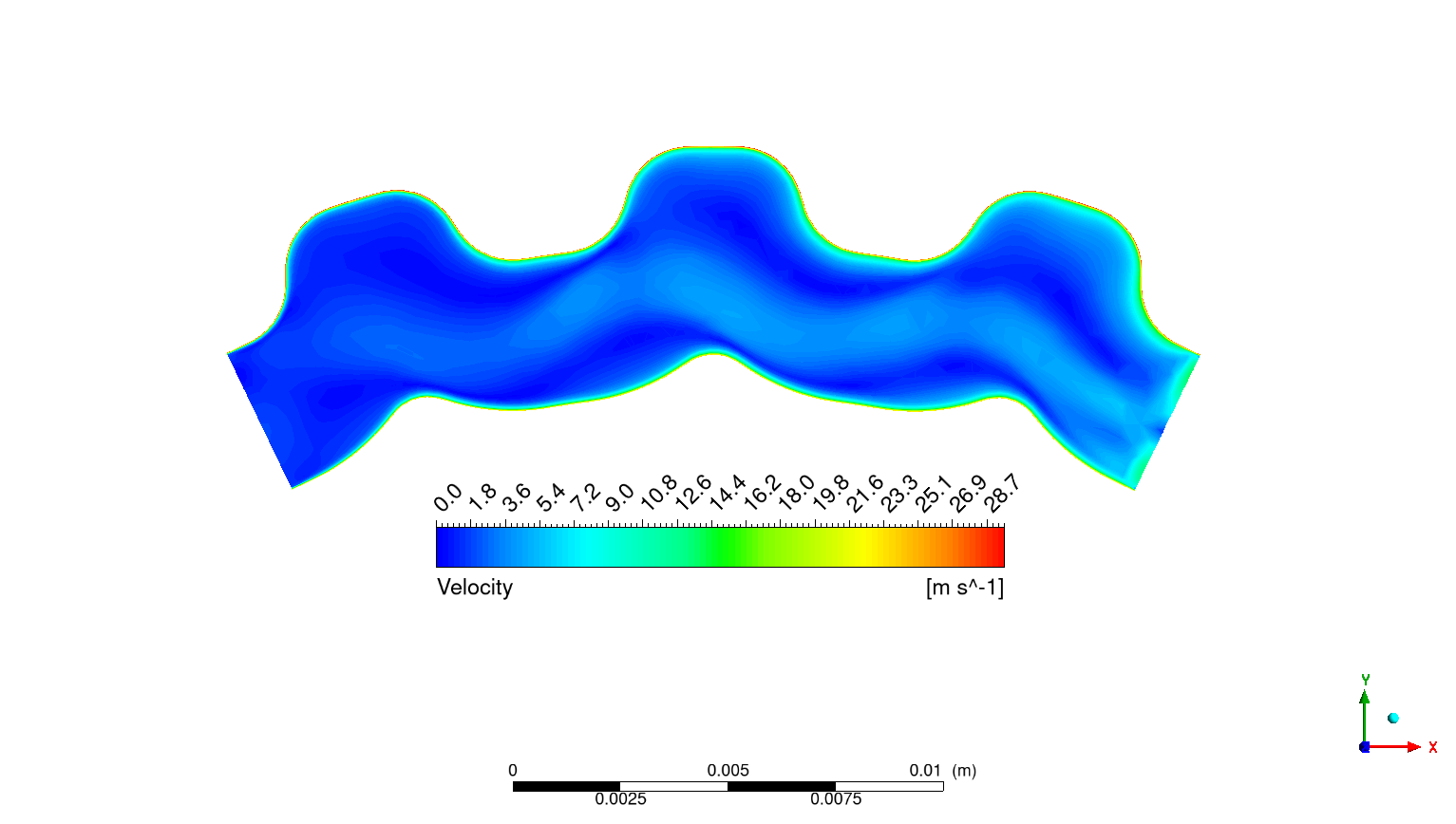}\label{fig:velocity-cross-model3-rpm-10000-intemp70}}
\hfill
\subfloat[T\textsubscript{in}=80\textdegree{}C]{\includegraphics[trim={5cm 6cm 5cm 3cm},clip,width=0.24\textwidth]{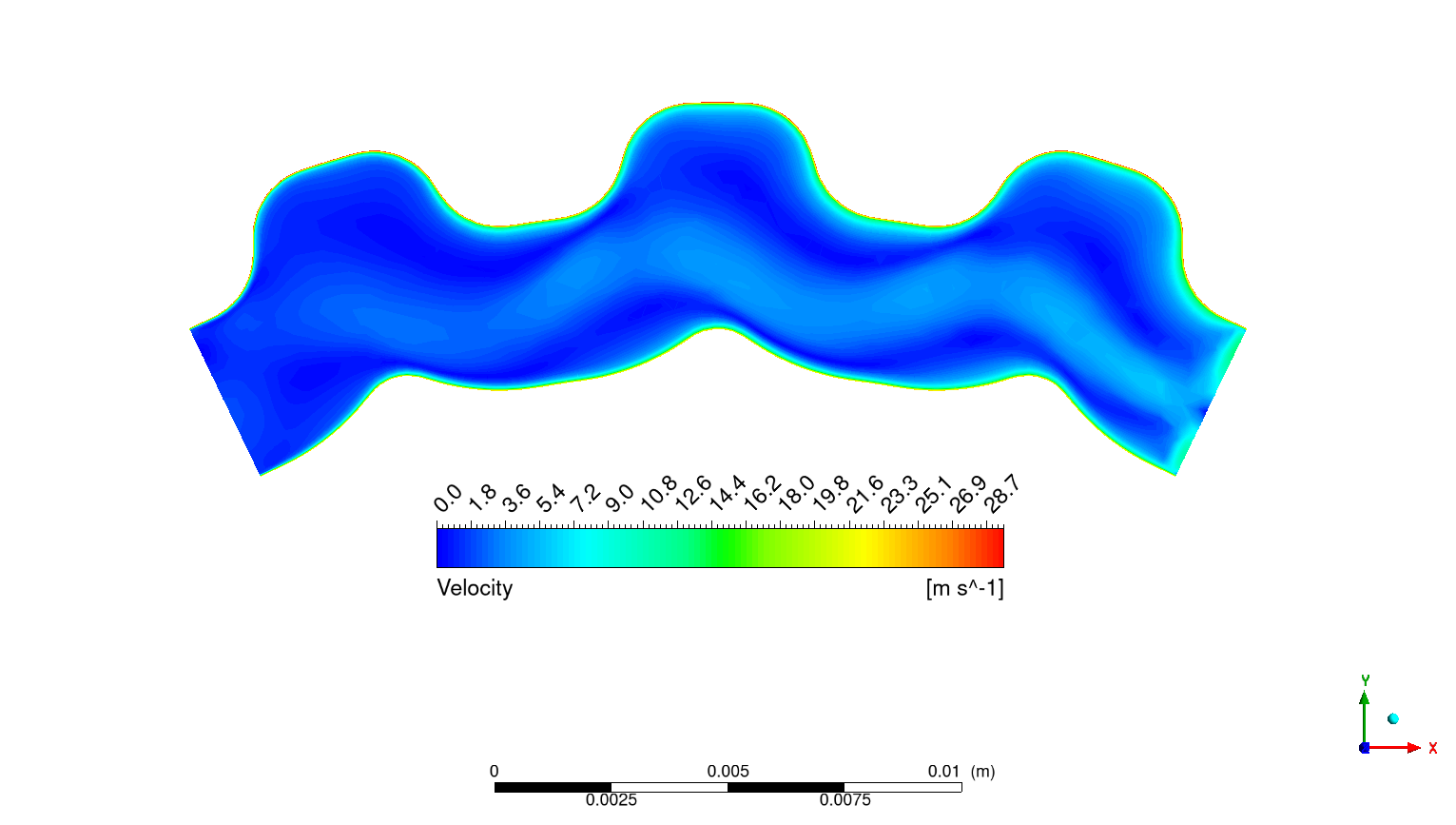}\label{fig:velocity-cross-model3-rpm-10000-intemp80}}
\caption{Cross-sectional view of velocity contours for Shaft Model 3 at 10,000 1/min across different inlet temperatures.}
\label{fig:velocity-cross-model3-intemp-comps-section3}
\end{figure}

In contrast, in Shaft Model~3, the velocity distribution within the cross-section changes more noticeably with rising inlet temperature (Fig.\ \ref{fig:velocity-cross-model3-intemp-comps-section3}). The wavy inner tube creates secondary flows and small recirculation zones. As the viscosity decreases, these recirculation zones expand or shift, and the local velocity peaks decline moderately. The design maintains sufficient turbulence and mixing to ensure effective heat transfer at the walls.

\begin{figure}[htbp]
\centering
\includegraphics[width=0.45\textwidth]{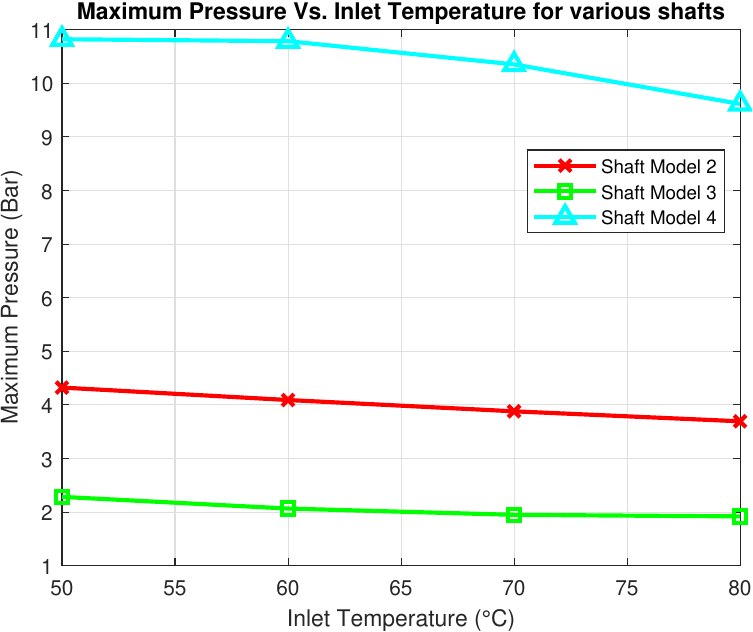}
\caption{Comparison of maximum pressure across all model shafts at various inlet temperatures.}
\label{fig:max-press-intemp}
\end{figure}

The reduced viscosity lowers frictional losses and thus the pump pressure. Higher inlet temperatures lower pressure most in Shaft Model 4, moderately in Shaft Model 2, and least in Shaft Model 3 (Fig.\ \ref{fig:max-press-intemp}). Under identical conditions, Shaft Model 3 records roughly half the pressure of Shaft Model 2. The wavy and more open channels in Shaft~3 restrict the flow less than the channels in Shaft Model 2. 

\begin{figure}[htbp]
\centering
\includegraphics[width=0.45\textwidth]{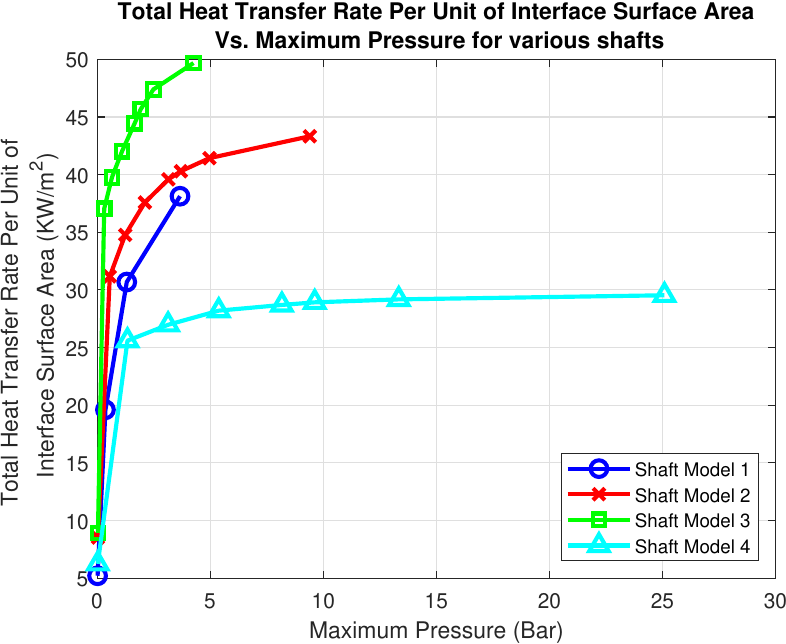}
\caption{Comparison of heat transfer rate per unit of interface surface area at various  maximum pressure.}
\label{fig:max-vel-intemp}
\end{figure}

Figure~\ref{fig:max-vel-intemp} illustrates the relationship between the total heat transfer rate per unit of interface surface area and the maximum pressure for the various shaft designs.  Shaft Model 3 outperforms the other designs as it achieves the highest heat transfer rates per surface at lower pressure levels. This increased heat transfer per necessary pressure results from the guidance of the coolant by the teeth. These teeth reduce the larger vortex streams, which cause the steep pressure increase in Shaft Model 1 but allow small vortices, which mix the coolant with less energy loss. The confinement of the ducts radially from the inside reduces the amount of liquid that does not contribute to cooling without increasing the pressure as much as in Shaft Models 2 and 4, which
Shaft Model 4 cannot translate its higher heat transfer due to the lager surface area (see Fig.\ \ref{fig:heat-inflowrate}) into an area-normalized advantage but lands last on the list in relative performance.

The rotor-cooling shaft concept with guided liquid in spoke-like channels outperforms the state of the art of a hollow cooling shaft by at least 30\% in heat transfer. Improvements through refinements allow improvements beyond 50\% (high speed) – 110\% (low speed) with approximately the same pressure level as a conventional hollow shaft and demonstrate the large potential of the concept. 

\section{Design and Manufacturing Aspects}

Effective rotor cooling in electric machines uses internal cooling channels that circulate coolant through the rotor shaft. This method absorbs heat generated within the rotor and reduces thermal resistance compared to external cooling.

\begin{figure}[htbp]
    \centering
    \begin{subfigure}{0.8\columnwidth} 
        \centering
        \includegraphics[width=\linewidth]{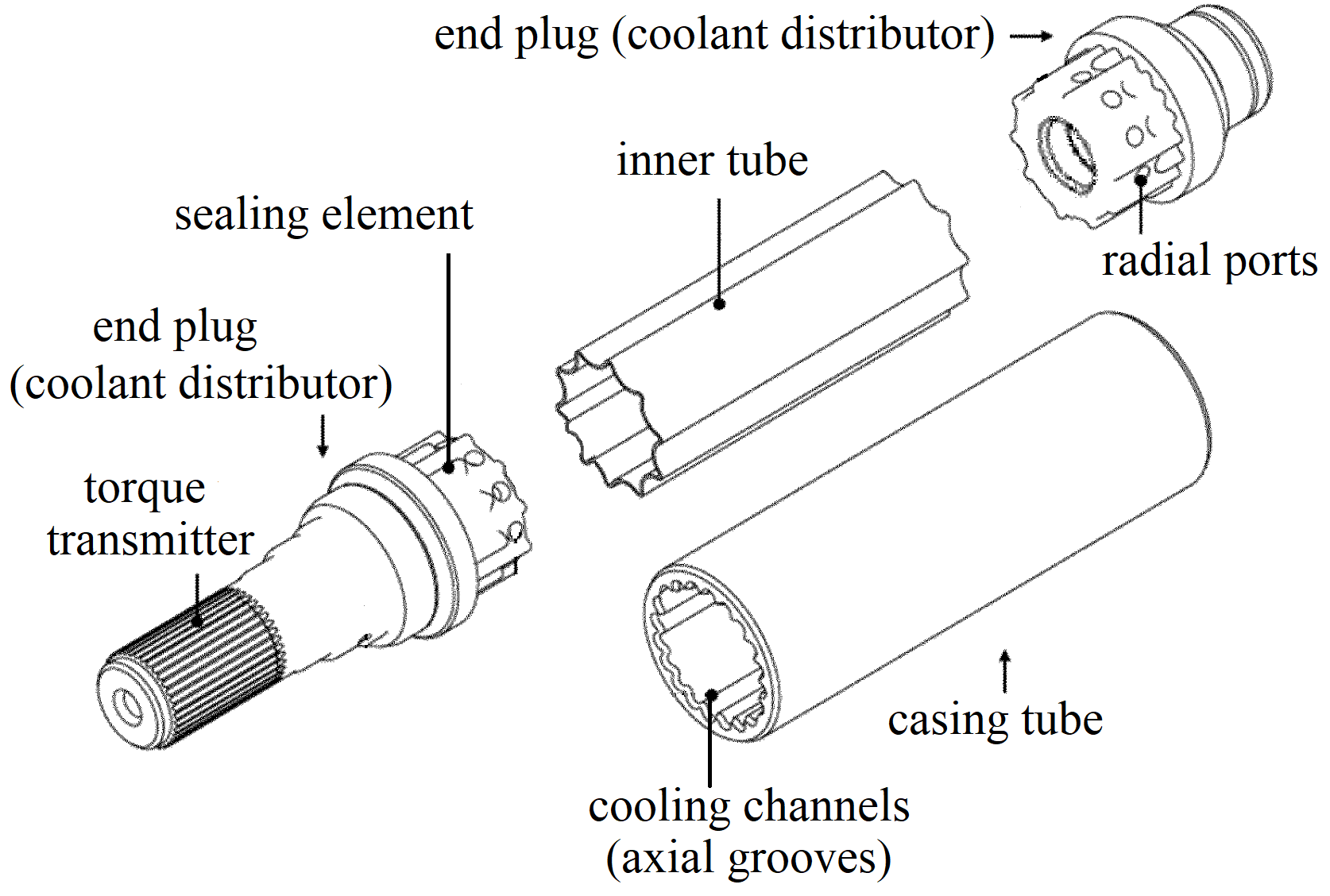}
        \caption{Exploded view of the rotor shaft assembly, showing the configuration of the outer rotor tube, inner tube, end plugs, and sealing interfaces.}
        \label{fig:Rotor-shaft_exploded_view}
    \end{subfigure}
    \hfill 
    \begin{subfigure}{0.8\columnwidth}
        \centering
        \includegraphics[width=\linewidth]{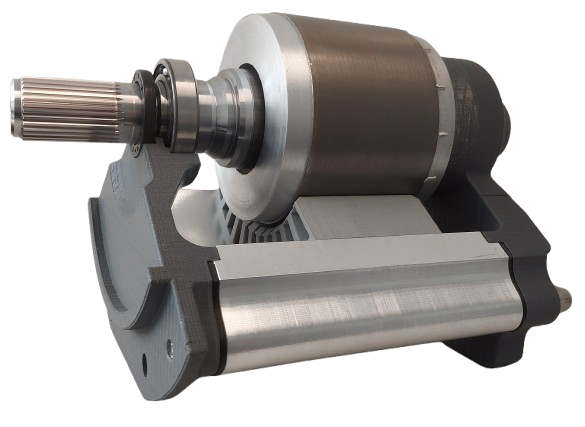}
        \caption{Implementation of the rotor shaft integrated into an induction-machine rotor.}
        \label{fig:quick2_motor}
    \end{subfigure}
    \caption{Overview of the rotor shaft design and its integration within the electric machine architecture.}
    \label{fig:patent_figs}
\end{figure}

The new shaft cooling concept integrates internal axial flow paths created by assembling a profiled rotor tube with matched end-pieces (Fig. \ref{fig:Rotor-shaft_exploded_view}). The assembly includes a casing tube, inner tube, and end connectors. The casing tube has an inner contour with axial grooves. These grooves enhance coolant circulation and increase the wetted internal surface area available for heat transfer. This design ensures uniform coolant distribution and consistent thermal management along the rotor's length.

The end plugs integrate multiple functions and provide structural integration, sealing, as well as fluid distribution. The machining ensures tight fits and reliable seals. The plugs withstand rotational forces and fluid pressures. Radial holes guide coolant into and out of the internal channels.

The rotor shaft components consist of 20MnCr5, a medium-carbon case-hardening steel. This material provides high toughness, strength, and wear resistance. Its favorable mechanical properties and thermal conductivity allow effective use of the cold-forming process in this design. Cold forming ensures precision, consistency, and enhances mechanical strength through induced work-hardening. This process achieves a balance between structural integrity and efficient thermal management.

The assembly process connects a profiled rotor tube with end plugs to create internal axial flow paths. The casing tube contains an inner profile with longitudinal grooves, while the inner tube may include a smooth, wavy, or matching profile. Positive and frictional engagement between these profiles defines the cooling channels and ensures reliable mechanical coupling. The interlocking profiles of the inner and outer tubes prevent unwanted relative motion and maintain structural stability during operation. Press-fit connection secures the end plugs to the casing tube; each plug engages a portion of the longitudinal grooves. Integrated cooling medium distribution structures within the end plugs guide coolant entry and exit efficiently. These structures channel the coolant through radial and axial passages to ensure thorough coverage of the cooling channels.

This manufacturing approach uses standard machining and forming techniques to maintain cost-effective and scalable production. The design avoids complex geometries and specialized tooling, which improves manufacturability and ensures precision. The modular structure enables precise quality control of individual components and simplifies maintenance and replacement. The straightforward design allows manufacturers to modify rotor shaft lengths, diameters, thicknesses, and profiles without requiring extensive adjustments to the manufacturing process.

The casing tube's internal profile features elevations and depressions along its circumference. These profiles can be periodically repeating, such as sinusoidal or helical patterns, to enhance coolant flow dynamics. The profile depth typically ranges from 1 to 6 mm, which ensures sufficient torsional strength and effective coolant distribution. The inner tube can also have a profile depth between 0.5 and 4 mm. 

Additionally, the casing tube may include tapered sections that accommodate rotational bearings, such as rolling bearings, within the electric machine (Fig. \ref{fig:quick2_motor}). This integrated design allows the shaft to transmit torque, manage cooling, and provide mechanical support.

These design and manufacturing choices establish a cohesive and practical approach to rotor cooling. The modular and scalable concept ensures efficient thermal management and supports various electric machine sizes and power ratings without introducing unnecessary complexity.

\section{Impact}
In induction machines, where rotor cooling is established (e.g., Tesla, Audi Q8 e-tron), this improvement can increase performance and reduce cost: (1) As induction machines are thermally limited by the rotor throughout most of their operating points even into field-weakening, improved rotor cooling immediately increases the motor power and power density of the otherwise very same motor design (until the stator winding turns into the limitation at some point typically for peak torque at low speeds when their electric loading determines the operation). Thus, the ability to cool 50\%–110\% more heat out of the rotor promises power increases of up to the same factor compared to rotor cooling with a conventional hollow shaft. (2) If operated at the same power, a motor with the improved cooling concept would reduce the temperature of the rotor bars. Due to the high temperature dependence of copper and aluminum, a switch from a hollow shaft to the improved version of the guided cooling ducts would cut the rotor loss by more than 20\%. (3) Alternatively, copper on the rotor could be cut by 20\% for equal efficiency. Although the improved cooling with the limited optimization performed so far not yet fully compensates a switch from copper to cheaper aluminum for same efficiency, it indeed does allow such a cost reduction if the same motor power is targeted. The rotor copper amounts to 40\%-50\% of the rotor material cost and about 15\%–20\% of the total material cost. The material (and in case of casting also manufacturing) cost reduction associated with a switch to aluminum promises a reduction of that share by at least a factor of two (at present even a factor of 4.2 due to the high copper cost since 2021).

For permanent-magnet synchronous machines, rotor cooling is particularly interesting for all operating points where the motor is rotor-critical. For most vehicle motors, that is the case in the higher speed range, which is reached on highways. Similar to induction machines, the improved cooling of the heat-sensitive rare-earth permanent magnets allows either cutting cost or increasing power. As magnets are practically the only temperature-sensitive material in the rotor and at present without liquid rotor cooling at all typically by far the limiting factor at higher speeds, a power increase of the permanent (S1) power on the highway of between 20\% and 40\%.

The current interest of European car manufacturers and suppliers to develop magnet-free field-wound synchronous machines (BMW, ZF, Mahle, Valeo, etc.) appears as an attractive application as it would eliminate one of the key bottle-necks. The gains in cooling exceed those of induction machines for two main reasons: (1) Field-wound machines have no liquid cooling at all to date, which limits the achievable excitation. (2) The magnet wire is substantially more temperature-sensitive and degradation-prone than the polymer-free bars in induction machines due to their insulating surface coating (e.g., Class H with 150\,°C average\,/\,180\,°C peak). Compared to current indirect cooling, optimized direct cooling promises with high certainty at least a factor of four, which would eliminate one major issue of field-wound motors that they cannot compete with rare-earth permanent magnet motors in torque (and power) density.

\section{Conclusion}
This paper presented a rotor shaft design with confinement and guidance of the liquid coolant through teeth near outer surface of the shaft. The shaft is designed such that the components can be manufactured through cold forming.
The confinement near the surface where the heat is handed over increased the heat transfer per necessary fluid pressure. The teeth increase the surface area and break the formation of larger vortices in the liquid, which tend to increase the necessary pressure, whereas small-scale vortices increase the heat transfer.

After establishing mesh independence and convergence, our analysis covered several parametric studies on the impact of rotational speed, inlet volumetric flow rate, and inlet temperature. 
Key findings were as follows:
\begin{enumerate}
    \item \textbf{Rotational Speed:} An increase in the shaft's rotational speed not only rises the outlet temperatures and heat transfer rates but also the fluid's pressure. Higher rotational speeds increase both the maximum pressure within the shafts and the fluid velocity.
    \item \textbf{Inlet Flow Rate:} Revealed that higher flow rates decrease outlet temperatures but increase the heat transfer rate, which describes a trade-off between cooling efficiency and the coolant's heat absorption capacity. This parameter also affected the pressure and velocity within the shafts, with higher flow rates slightly increasing the system's overall pressure and maintaining relatively stable velocity profiles.
    \item \textbf{Inlet Temperature:} Elevating the inlet temperature consistently resulted in higher outlet temperatures across all models.
\end{enumerate}

Among the designs, \textit{Shaft Model 3} consistently demonstrated superior performance. It excelled in optimizing the balance between cooling efficiency and pressure management, attributed to its unique design that facilitates effective fluid mixing and heat dissipation.

In conclusion, this CFD analysis provided insightful revelations on the performance dynamics of different shaft designs under varied operational conditions. The study not only highlights the critical role of design elements in thermal management efficiency but also lays a groundwork for future enhancements. Moving forward, focusing on innovative design features that promote better fluid dynamics and thermal interactions, like those observed in Shaft Model 3, could pave the way for developing more efficient cooling systems. Further research could explore additional design modifications, alternative materials, and simulation methods to amplify these outcomes.

\bibliographystyle{IEEEtran}

\end{document}